# Mechanically Assisted Symmetry Reconstruction for Extraordinary Piezoelectricity

Jinhui Fan[1]†, Chonghe Wang[2]†, Xiaoyan Lu[1]*, Yunpeng Ma[3], Zijian Hong[4], Yuzhao Qi[1], Yanzhe Dong[1], Xiaoyue Zhang[5], Chuchu Yang[6], Yongchun Zou[7], Xu Zheng[8], Xiaolong Li[8], Qian Li[2], Xiang Xu[1], Si-Young Choi[9], Jiyan Dai[10], Wenwu Cao[11], Dragan Damjanovic[12,13]*, Hui Li[1]*

[1] Department of Mechanics, School of Civil Engineering, Harbin Institute of Technology; Harbin, 150001, China.

[2] School of Mechanics and Engineering Science, Peking University; Beijing, 100871, China.

[3] State Key Laboratory of New Ceramic Materials, School of Materials Science and Engineering, Tsinghua University; Beijing, 100084, China.

[4] State Key Laboratory of Silicon and Advanced Semiconductor Materials, School of Materials Science and Engineering, Zhejiang University; Hangzhou, 310027, China.

[5] Guangdong Provincial Key Laboratory of Magnetoelectric Physics and Devices, School of Physics, Sun Yat-sen University; Guangzhou, 510275, China.

[6] Center for Composite Materials and Structures, Harbin Institute of Technology; Harbin, 150001, China.

[7] Center for the Analysis and Measurement, Harbin Institute of Technology; Harbin, 150001, China.

[8] Shanghai Synchrotron Radiation Facility, Shanghai Institute of Applied Physics, Chinese Academy of Sciences; Shanghai, 201204, China.

[9] Department of Materials Science & Engineering, Pohang University of Science and Technology; Pohang, 37673, Korea.

[10] Department of Applied Physics, The Hong Kong Polytechnic University; Hong Kong, China.

[11] Center of Acoustic Functional Materials and Applications, School of Materials Science and Intelligent Engineering, Nanjing University; Suzhou, 215163, China.

[12] School of Physics, Harbin Institute of Technology; Harbin, 150001, China.

[13] Institute of Materials, Swiss Federal Institute of Technology–EPFL; Lausanne, 1015, Switzerland.

*Corresponding author. Email: luxy@hit.edu.cn; dragan.damjanovic@epfl.ch, lihui@hit.edu.cn.

†These authors contributed equally to this work.

**Abstract:** Active symmetry control—a central challenge in materials science, particularly in ferroelectrics—is achieved via mechanically assisted poling (MAP) guided by thermodynamics and phase-field modeling. This approach yields extraordinary piezoelectric coefficients (~5,000 $pC \cdot N^{-1}$ at 24°C; 11,700 $pC \cdot N^{-1}$ at 58°C) together with ~65% optical transmittance in a classic relaxor ferroelectric, $Pb(Mg_{1/3}Nb_{2/3})O_3$-$PbTiO_3$. Mechanical suppression of undesirable phases stabilizes a reconstructed symmetry with highly ordered domains, verified by multiple characterization techniques. The strategy is validated across several distinct ferroelectric systems. To demonstrate its practical utility, we fabricate a transparent dual-modal wearable sensor integrating continuous blood pressure monitoring via piezoelectricity with photoplethysmographic SpO2 detection, enabling high-fidelity physiological tracking. This work establishes mechanically assisted symmetry reconstruction as a pathway to multifunctional optoelectronic materials and compact wearable health technologies.

## Introduction

Active and precise control over crystal symmetry has long stood as a fundamental challenge in materials science, as it directly governs material properties from piezoelectricity to optical transparency and beyond (*1-3*). This challenge is particularly acute in ferroelectric single crystals since once crystallized, piezoelectricity is hindered by fixed phase symmetry and domain configurations (*4,5*). To date, efforts to enhance piezoelectric response have largely focused on flattening the free-energy landscape, primarily through the compositionally driven morphotropic phase boundaries (MPBs) (*6-18*), domain engineering (*19,20*), and local heterogeneity (*21-23*). These modifications inevitably generate polymorphic microstructures and internal stress/strain, stabilizing low-symmetry phases that account for the enhanced piezoelectricity (*24-26*). External mechanical force has been explored as an alternative to internal stress/strain for domain configurations during poling, thereby avoiding the high-field breakdown of the crystals (*27,28*). Yet, despite a 40% reduction in poling field, this approach has yielded only a modest ~10% enhancement in piezoelectric response (*29,30*). Recent innovations, such as alternating current poling (ACP), have advanced the field further, achieving 65% optical transmittance while retaining a high piezoelectric response ($d_{33}$ ~ 2,100 $pC \cdot N^{-1}$) in rhombohedral PMN-0.28PT (*3*). Despite these advances, this method encounters a fundamental limitation at MPB: coexisting phases and ensuing light scattering induce the inherent trade-off between piezoelectricity and optical transparency (*31,32*). In addition, the presence of lower-performance phases (*33,34*) also cap the ultimate performance. Thus, rather than merely accommodating complex disordered states, active control of crystal symmetry with a selected phase is the key to break the current performance ceiling and unlock new regimes of multifunctional integration for transformative applications ranging from opto-electro-mechanical systems to monolithic wearable devices (*35–37*).

Herein, we introduce mechanical field as a powerful and previously overlooked dimension for symmetry reconstruction. A mechanically assisted poling (MAP) approach, using a tailored mechanical force applied during electric poling, yielded ultrahigh piezoelectric properties ($d_{33}$ > 5,000 $pC \cdot N^{-1}$) with near-perfect transmittance (~ 65% for visible light; see Methods) in PMN-0.33PT with MPB composition. Such significant piezoelectric enhancement was attributed to MAP-induced symmetry reconstruction. This process operates at the critical rhombohedral-orthorhombic (R-O) phase boundary with a flat energy well, enabling the conversion of multiphase coexistence into monophase domains. By suppressing undesirable intermediate phases, a pure, highly ordered rhombohedral state was stabilized, unlocking a previously inaccessible combination of a record-high piezoelectric coefficient and exceptional optical transparency.

Remarkably, the principle of symmetry reconstruction by MAP shows broad applicability beyond the PMN-PT system. We demonstrate its efficacy in reconstructing symmetry across diverse ferroelectric single crystals, including classical single crystal $BaTiO_3$, binary relaxor ferroelectric $Pb(Zn_{1/3}Nb_{2/3})O_3$-$0.065PbTiO_3$ (PZN-0.065PT), and ternary $Pb(In_{1/2}Nb_{1/2})O_3$-$0.38Pb(Mg_{1/3}Nb_{2/3})O_3$-$0.33PbTiO_3$ (PIN-0.38PMN-0.33PT) crystals. This general paradigm establishes a new design rule in which coupled electric and mechanical fields are used to actively drive materials into desired, thermodynamically inaccessible states through controlled polarization-rotation paths, thereby enabling next-generation materials with on-demand functionalities. Crucially, the high piezoelectricity and optical transparency achieved simultaneously directly solve a key integration barrier in wearable sensing. We demonstrate this with a single transparent device that unifies blood pressure and $SpO_2$ monitoring—a dual-modal capability unattainable with opaque piezoelectric materials. This application prototype underscores how mastering crystal symmetry can redefine functionality for compact, multifunctional health monitors and optoelectronic systems.

## Principles for a Targeted Phase Architecture

Pushing the limits of piezoelectricity in ferroelectrics involves navigating the flattened free-energy landscape at the MPB, particularly suppressing the undesirable intermediate phases with lower performance. This task, however, presents a significant challenge due to the minimal free energy difference ($\Delta G < 10^3$ J·m$^{-3}$) between the coexisting phases, which prevents effective selection of designed phases via conventional processing. Understanding this intricate thermodynamic landscape is therefore essential to reconstruct a phase-pure, high-performance crystal structure.

Within the Landau-Ginzburg framework, the total free energy ($F_{\mathrm{total}}$) is expressed as the sum of Landau ($F_{\mathrm{Land}}$), elastic ($F_{\mathrm{elas}}$), electrostatic ($F_{\mathrm{elec}}$), and gradient ($F_{\mathrm{grad}}$) contributions (*38,39*). Specifically, intrinsic equilibrium phases correspond to local minima of the Landau energy $F_{\mathrm{Land}}$ (*34*). A key feature of the MPB is the softening of the free-energy landscape, where the curvature near the minima becomes shallow. Mathematically, this softening is captured by the near-zero eigenvalues of Hessian matrix $H_{ij} = \partial^2 F_{\mathrm{Land}}/\partial P_i \partial P_j$ with degenerated polarization components $P_i$. These low-energy barriers facilitate easy polarization rotation, leading to the enhanced dielectric response ($\varepsilon_{ij}^{-1} = H_{ij}$) and giant piezoelectricity via the constitutive relation: $d_{ijk} = 2\varepsilon_0 \varepsilon_{im} Q_{mjkl} P_l$ ($i, j, k, m, l$ = 1, 2, 3), where $Q_{mjkl}$ is the electrostrictive tensor. In particular, the longitudinal piezoelectric coefficient $d_{33}$ (in Voigt notation) governs the sensitivity in longitudinal vibration mode, a primary working mode for bulk transducers, sensors and actuators.

The piezoelectric coefficient is highly sensitive to the crystal orientation, as evidenced by the superior performance of the R phase along pseudocubic $[001]_{\mathrm{pc}}$ direction (e.g. $d_{33}^{\mathrm{R}} \approx 2{,}500$ pC·N$^{-1}$ in PMN-PT) compared to other orientations (*40,41*). This strong dependency makes the strategic selection of a target phase and orientation critical for achieving superior performance. In PMN-$x$PT's MPB compositions ($x$ = 0.31–0.35), this task is both promising and challenging due to the complex phase coexistence, which produces a flattened Landau energy landscape (Fig. S1). At critical composition of $x$ = 0.33, the Landau free energies for R and O phases are nearly equal (Fig. 1a). Slight modifications of the PT content or perturbations by external fields can disrupt this delicate balance, favoring either the O phase with *mm*2 symmetry (Fig. 1b) or the R phase with 3*m* symmetry (Fig. 1c). According to the Landau-Devonshire model calculations, the piezoelectricity of the R phase with flat energy well is higher than that of O phase with steep energy well in PMN-$x$PT (Fig. S2). Therefore, we focus on phase engineering near the critical composition PMN-0.33PT to maximize the piezoelectricity. Our theoretical predictions show that the piezoelectric coefficient $d_{33}$ of the R phase along $[001]_{\mathrm{pc}}$ direction can exceed 5,000 pC·N$^{-1}$, considerably higher than that of the O phase (~1,600 pC·N$^{-1}$, Fig. S3). Thus, stabilizing the R phase in a $[001]_{\mathrm{pc}}$-orientated crystal with a PT content around 0.33 is essential.

Concurrently, optical transparency is governed primarily by domain structures which can be optimized with polarization vectors aligned into specific configurations along the light propagation direction to reduce the light scattering at the domain walls. Conventional direct current poling (DCP) along $[001]_{\mathrm{pc}}$ results in single tetragonal phase structure (low $d_{33}$), 4R (71° and 109° domain walls), or 4O (60° and 90° domain walls) structures (Fig. 1d). While 4R and 4O structures are opaque due to the overlap of refractive index ellipsoids of different do这个 main variants (Fig.

1d), 2R and 2O structures offer enhanced transparency through non-overlapping optical projections (Fig. 1e).

Therefore, to simultaneously achieve ultrahigh piezoelectricity and excellent transparency, it is essential to efficiently stabilize a pure, highly ordered state of the high-performance R phase while completely suppressing the formation of the lower-performance O and T phases. To achieve a structure with superior piezoelectricity and optical transparency, we propose a MAP strategy to reconstruct the crystal symmetry by stabilizing 2R (109° domain wall) architectures from initially R-O phase-coexisting PMN-0.33PT single crystals (Fig. 1f).

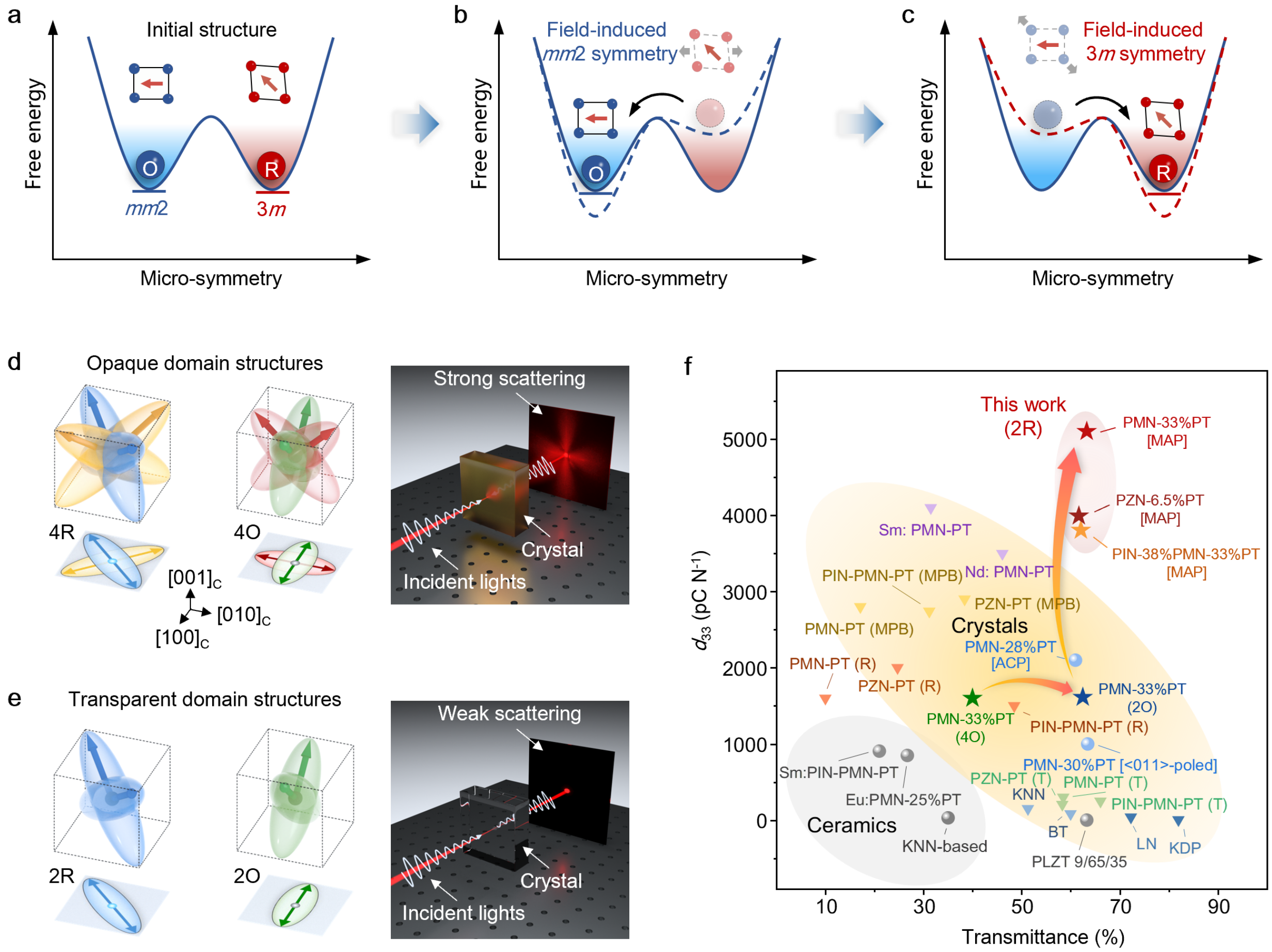

**Fig. 1 Symmetry reconstruction enabling high-performance transparent piezoelectric materials. a, Illustration of free energy profile for R-O phase coexisting system. b, c,** MAP enabled symmetry reconstruction of O phase with *mm*2 symmetry and R phase with 3*m* symmetry as dominating state by shifting the energy profile. The gray arrows show direction of the applied external field relative to the initial polarization with direction along the red arrows. **d,** Schematic of the optical indicatrix ellipsoids for 4R and 4O domain structures, and their crossed projections causing strong light scattering at domain boundaries. **e,** Optical indicatrices of 2R and 2O domains are aligned $(001)_{pc}$ with minimized light scattering. **f,** Piezoelectric coefficients and transparency of selected ferroelectric materials (see Table S1). Our results of the MAP materials, marked by the stars, achieve a piezoelectric coefficient of up to 5,100 pC·N$^{-1}$ at ambient temperature, representing a twofold increase over conventional DC-poled 4O structures, while also maintaining

exceptional optical transparency. Phase structure (R, T, MPB) and poling method are given in parentheses and square brackets, respectively.

## Mechanically Assisted Symmetry Reconstruction

To suppress the O phase with inferior piezoelectricity, and stabilize the R phase with 2R domain structures, the energy profile must be carefully tailored to enable favorable polarization rotation paths for the R phase stabilization. To visualize the energy landscape of coexisting intrinsic phases, we map the free energy $G$ considering only $F_{\mathrm{Land}}$ of PMN-0.33PT in spherical coordinates, $G(|P_s|, \theta, \varphi)$. Without external fields, the system resides in a thermodynamically stable '12O + 8R' multiphase configuration, characterized by a flattened energy barrier between near-degenerate potential wells (Fig. S4). Stabilizing the 2R domain architecture (e.g., polarization aligns within $\{110\}_{pc}$ plane) requires anisotropic field to reshape the energy landscape. We show that a critical strategy involves concurrent application of electric fields and mechanical force (e.g., uniaxial tension or compression) to reorient polarization vectors along diagonal crystallographic directions. In this work, we adopt mechanical compression using a standard uniaxial loading setup.

To investigate the stabilization of R phase from the R-O coexisting phases in PMN-0.33PT under both mechanical and electrical fields, we calculated the map of polarization modulus $|P_s|$ using Landau-Devonshire theory with initial state of the O phase (Fig. 2a). As illustrated in the phase map, the O phase maintains its stability with minimal distortion under $[001]_{pc}$ electric field until transforming into the T phase. Interestingly, application of mechanical force along the diagonal $[1\bar{1}0]_{pc}$ direction promotes the formation of R phase. The stability of the MAP-induced 2R state was further verified by our phase equilibrium simulation (Fig. S5). The simulations proceeded with the following steps: (1) DCP along the $[001]_{pc}$ direction to align the polarization variants upward; (2) Phase selection via subsequent mechanical compression along the $[1\bar{1}0]_{pc}$ direction to form a 2R architecture; (3) Removal of the electric field; (4) Removal of mechanical force. All identified phases were validated through the Hessian matrix analysis. Calculation results reveal that the initial R-O coexistence state transforms into a distorted O phase under a $[001]_{pc}$-oriented DCP field ($E$ = 10 kV·cm$^{-1}$), and subsequently into a 2R structure with minor distortion under the uniaxial compression ($\sigma$ = 100 MPa). After removal of all external fields, the 2R structure is stabilized, demonstrating the feasibility of targeted polarization distribution manipulation via the MAP strategy.

To facilitate the application of a DC field along $[001]_{pc}$ direction and a mechanical force along the $[1\bar{1}0]_{pc}$ diagonal direction, we cut the $[001]_{pc}$ -oriented PMN-$x$PT plates into rectangular samples with dimensions of 4.2 mm//$[110]_{pc}$ × 3.2 mm//$[1\bar{1}0]_{pc}$ × 0.5 mm//$[001]_{pc}$. Then, we applied a DC electric field of 14 kV·cm$^{-1}$ (4 kV·cm$^{-1}$ higher than the theoretical value to ensure effective poling) and a compressive stress of 100 MPa (corresponding to a force of 160 N over a 0.5 × 3.2 mm$^2$ area) (Fig. 2b). Polarized light microscopy (PLM) image revealed that the region under applied force transformed into the R phase, in contrast to the area without force, which remained as the O phase (Fig. 2c). This transformation indicates the successful symmetry reconstructed of the 2R structure, consistent with phase-field simulations (Fig. S6). The resulting 2R architecture exhibits exceptional piezoelectricity and optical transparency: large $d_{33}$ coefficient increased by over 200%, rising from 1,600 pC·N$^{-1}$ up to 5,100 pC·N$^{-1}$ (Fig. 2d), and visible-light transmittance along $[001]_{pc}$ increased from ~40% (O state) to over 60% (2R state) (Fig. 2e).

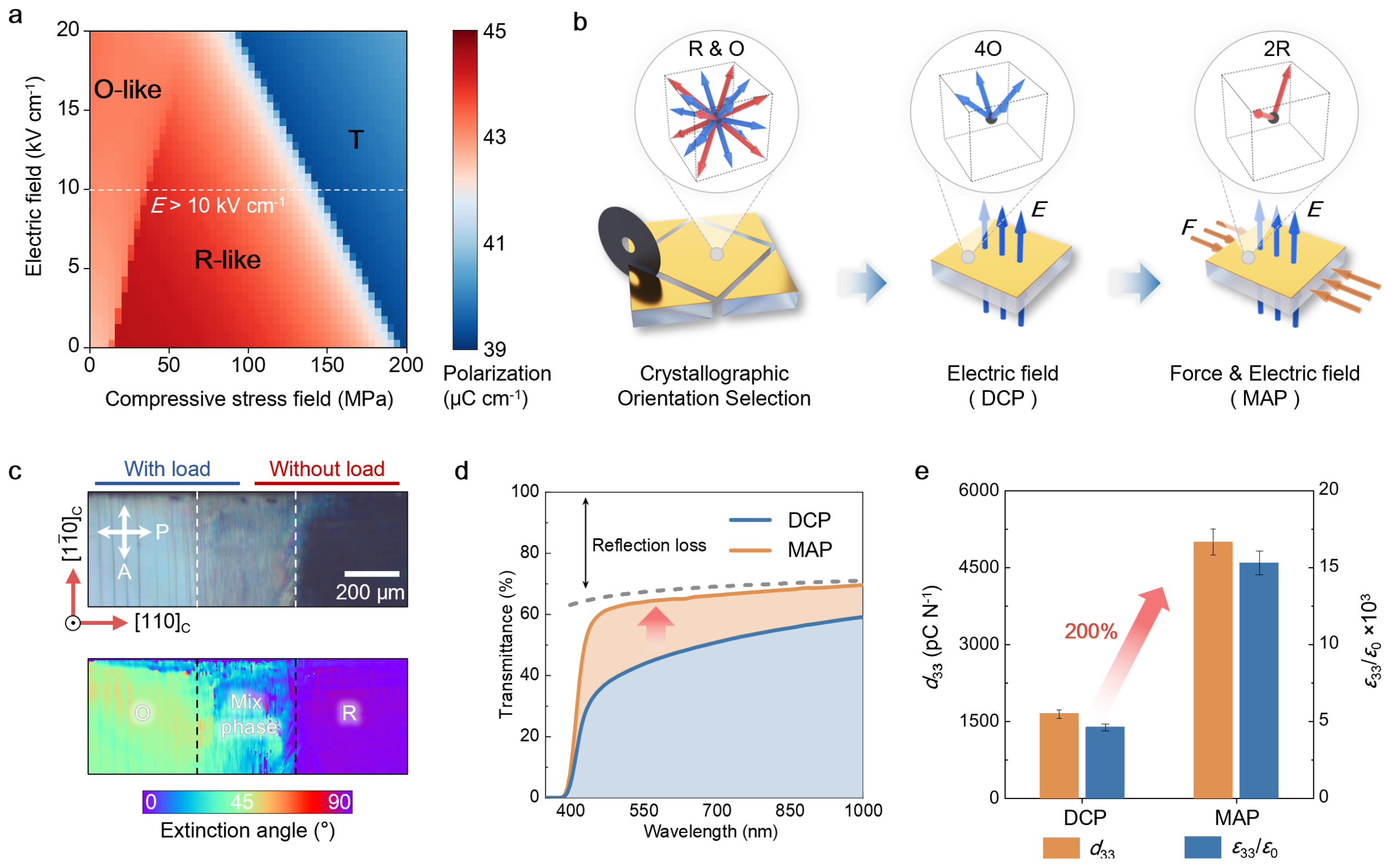

**Fig. 2 Stabilization of R phase via MAP for symmetry reconstruction and enhanced piezoelectric performance. a,** Phase diagram of PMN-0.33PT crystals. **b,** Experimental procedure for achieving the 2R phase structure using the MAP approach in PMN-0.33PT crystals. **c,** PLM images of domain structures with and without force. **d,** Comparison of the piezoelectric coefficient $d_{33}$ (orange bars) and dielectric constant $\varepsilon_{33}/\varepsilon_0$ (blue bars) between DCP- and MAP-processed samples. **e,** Comparison of optical transmittance spectra.

## Structural Characterization

The symmetry reconstruction in PMN-0.33PT crystals was examined using 3-D high-resolution reciprocal space mapping (RSM) to reveal the differences in domain structure (Methods). Conventional DCP-poled samples exhibit three diffuse diffraction peaks near the $(202)_{pc}$ Bragg reflections (Fig. 3a). The three peaks separate on the $(\bar{1}10)$ plane, indicating the presence of four O variants, with the $O_2$ and $O_4$ domains overlapping into a single peak along the $(202)_{pc}$ direction (Fig. 3b). The lattice spacing for the $O_1$ domain is larger than that of $O_3$, resulting in a higher reciprocal lattice vector for the $O_1$ with $\Delta Q_L = 0.019$ (Fig. 3c, d). In contrast, MAP-processed samples show a uniform distribution of reflections with two distinct, less diffuse Bragg spots and smaller peak separation ($\Delta Q_L = 0.014$) on the $(\bar{1}10)$ plane (Fig. 3e–h). Further RSM analysis of the $(222)_{pc}$ and $(\bar{2}22)_{pc}$ Bragg reflections in MAP samples confirms the successful stabilization of $R_1$ and $R_3$ domain configurations (Fig. S7). This 2R state exhibits exceptional optical homogeneity along the $[001]_{pc}$ direction, as confirmed by polarized light microscopy (PLM) observations (Fig. S8).

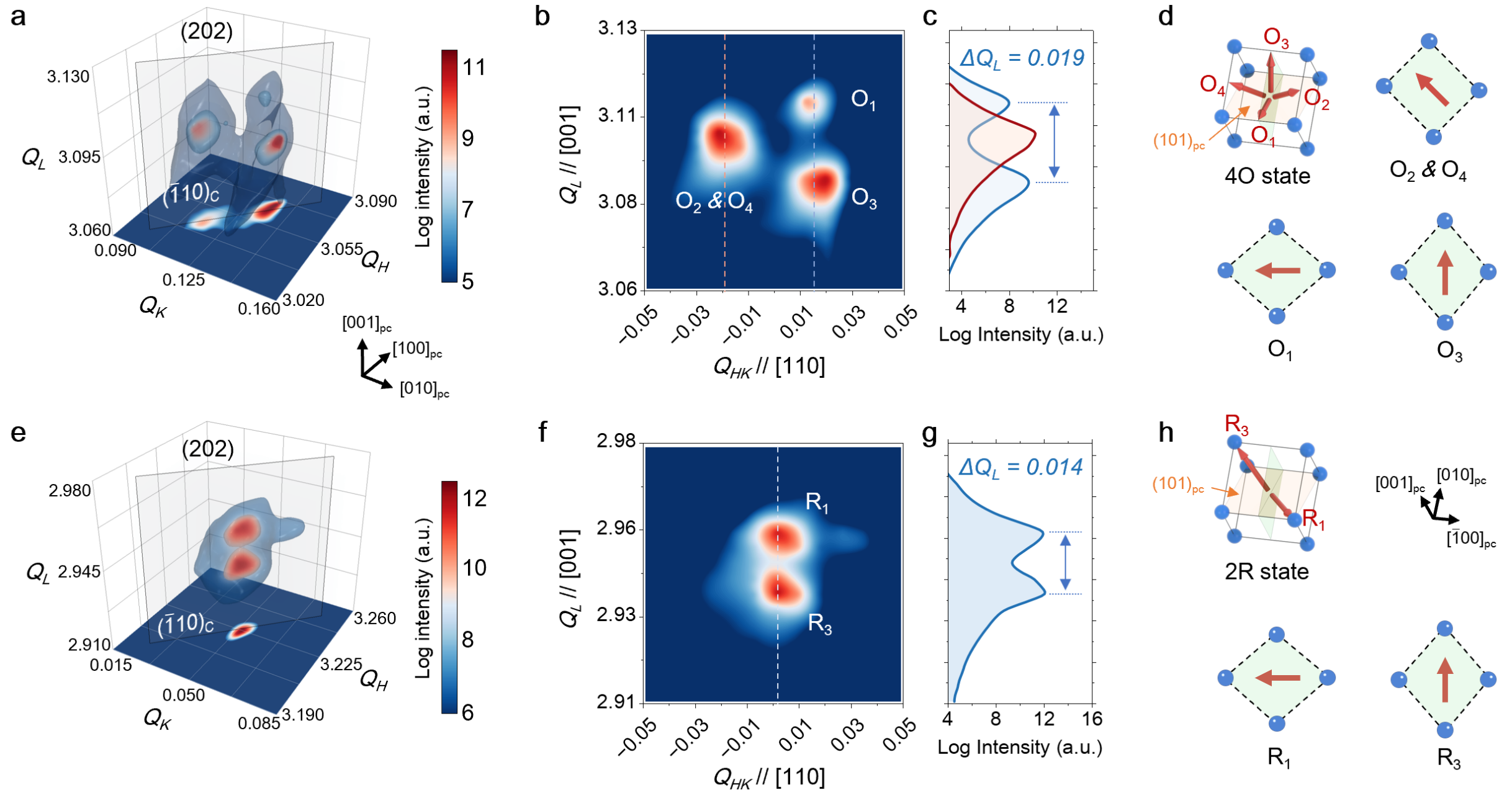

**Fig. 3 RSM of [001]$_{pc}$-oriented PMN-0.33PT crystals processed by DCP and MAP methods. a,** 3D-RSM of $(202)_{pc}$ reflection for the DCP-processed samples, showing three diffraction spots corresponding to O variants. **b,** In-plane RSM on the on ($HK$0) plane reveals three coplanar diffraction spots on $(\bar{1}10)_{pc}$ plane (left), associated with four polarization variants (from $O_1$ to $O_4$). Extracted peak positions of the variants (right), where $O_1$ and $O_3$ domains exhibit a notable lattice vector difference of $\Delta Q_L = 0.019$. **c,** Schematic illustration of the 4O variants, featuring lattice distortions along three orientations. **d,** 3D-RSM of the $(202)_{pc}$ reflection for the MAP-processed samples, showing only two diffraction spots. **e,** In-plane RSM on the $(\bar{1}10)_{pc}$ plane (left), and extracted peak positions of the $R_1$ and $R_3$ variants with lattice vector difference of $\Delta Q_L = 0.014$ (right). **f,** Schematic of 2R variants with lattice distortions along diagonal directions.

## Polarization Characterization and Piezoelectric Measurements

To further investigate the domain structures and their relationship with piezoelectricity, a second-harmonic generation (SHG) microscopy was employed to examine the mesoscale domain variants. Unlike the nearly straight 90° domain walls separating $O_1/O_3$ or $O_2/O_4$ variants observed in DCP-processed samples (Fig. 4a), MAP-processed crystals develop large R-109° domains on the $(110)_{pc}$ plane (Fig. 4b). Compared to the domain size of DCP-processed samples (average ~1.5 μm, Fig. 4c), the MAP-processed crystals display meandering domains with an average size of up to ~15 μm (Fig. 4d), as identified via SHG microscopy. Additionally, the presence of notable SHG intensity fluctuation in the MAP-processed samples further indicate an easy polarization shift from intrinsic R phase to R-like monoclinic phases along R-O rotation path (Fig. S9). Dielectric spectra offer additional evidence for the facile polarization rotation, showing enhanced dielectric permittivity in MAP-processed samples relative to their DCP-processed counterparts (Fig. 4e). Dielectric spectroscopy also confirms the thermodynamic stabilization of MAP-processed domain structure via a distinct R-O transition peak at 58 °C, further supported by PLM observations (Fig. S10). A remarkable electric-field-induced strain of -0.027% is measured at 26 °C via laser interferometry ($E$ = -0.5 kV·cm$^{-1}$, 10 Hz), representing a 195% enhancement over DCP-processed counterparts. The corresponding piezoelectric coefficient rises from ~5,000 pm·V$^{-1}$ (24 °C) to 11,700 pm·V⁻¹ (58 °C), peaking near the critical temperature – consistent with a first-order R-O phase transition (Fig. 4f).

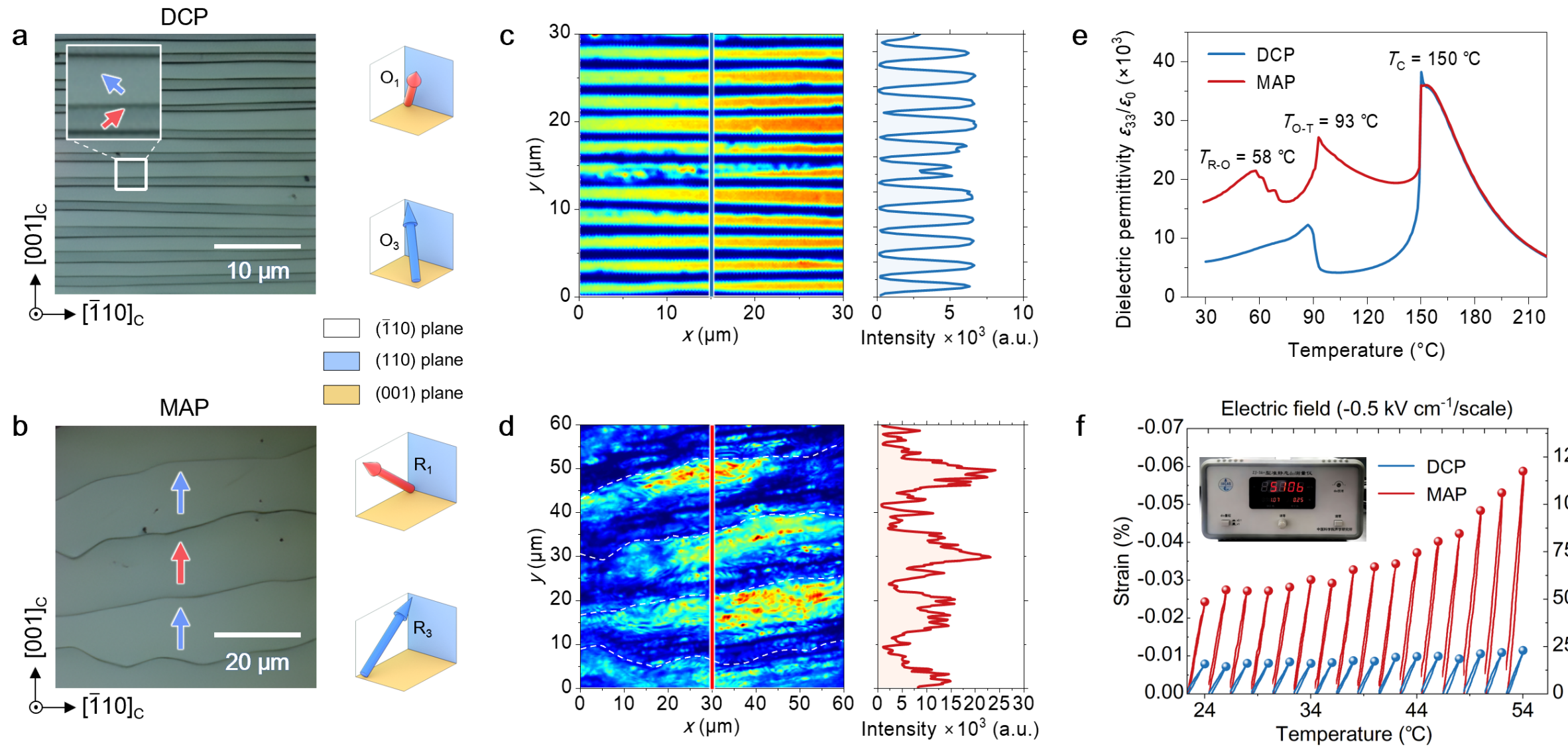

**Fig. 4 Domain structures and their correlation with dielectric and piezoelectric responses of PMN-0.33PT crystals processed by DCP and MAP methods. a, b,** Optical images of chemically etched $(110)_{pc}$ surfaces of DCP-processed sample with dense stripe domains (~1.5 μm) and MAP-processed sample with sparse, broader domains (~15 μm), respectively. **c, d,** SHG mapping corresponding to panels a and b, respectively. The SHG intensity of MAP-processed sample shows strong fluctuation, indicative of increased local structural heterogeneity. **e,** Temperature-dependent dielectric permittivity $\varepsilon_{33}/\varepsilon_0$ and loss factor measured at 1 kHz. The MAP-processed sample displays three distinct ferroelectric phase transitions, whereas only two are observed in the DCP-processed sample. **f.** Electric-field-induced strain measured at various temperatures under a driving field of -0.5 kV cm$^{-1}$ at 10 Hz. MAP-processed sample exhibits enhanced strain responses compared to its DCP-processed counterpart with $d_{33}$ ~5,000 pm·V⁻¹ at room temperature.

## Validation

By applying anisotropic mechanical and electric fields, MAP reconstructs crystal symmetry to produce unique domain architectures. This approach has proven to be particularly effective for typical R-phase PMN-$x$PT systems, especially in $[001]_{pc}$-oriented PMN-0.31PTwithout the crystal damage or cracking. We could process crystals of this composition in plates as large as 20 × 20 mm$^2$, the largest transparent PMN-PT single crystal yet reported (Fig. S11). Furthermore, the generality of this method was verified across diverse ferroelectric single crystals including classical single crystal $BaTiO_3$, binary relaxor ferroelectric PZN-0.065PT, and ternary PIN-0.38PMN-0.33PT crystals (Fig. S12).

## Dual-Modal sensing demonstration

To directly validate the unique potential enabled by this combination of the piezoelectricity and optical transparency, we fabricated a dual-modal wearable physiological sensor to simultaneously and independently acquire blood pressure (BP) waveform via transparent ultrasound transduction (TUT) and peripheral oxygen saturation ($SpO_2$) via photoplethysmography (PPG) (Fig. 5a). Crucially, the optical transparency of the MAP-based TUT enables $SpO_2$ sensing by allowing essential light transmission, in sharp contrast to the DCP-based one (Fig. 5b). The MAP-based TUT has a large peak-to-peak voltage ($V_{pp}$ = 3500 mV under 1.55 μJ electric pulse excitation) and a wide -6 dB bandwidth (~71%) with center frequency ~8 MHz, much superior to that of the DCP-based TUT with $V_{pp}$ of 900 mV and -6 dB bandwidth of 56% (Fig. 5c, 5d). Acoustic field simulation of our TUT (element size: 3 × 3 mm) demonstrates a well-collimated beam with a −6 dB lateral width of ~1.0 mm and sustained acoustic penetration beyond 4 cm into tissue (Fig. 5e). Results of continuous monitoring on a healthy adult subject after moderate-intensity exercise showed an elevated systolic blood pressure of approximately 150 mmHg and a peripheral oxygen saturation ($SpO_2$) of ~94% in the immediate post-exercise period. (Fig. 5f). Following a 2-minute rest, BP decreased to around 99 mmHg, while $SpO_2$ recovered to 97% (Fig. 5g). These trends are fully consistent with the expected post-exercise physiology (declining cardiac output and restored gas exchange), confirming the sensor's functionality in tracking dynamic physiological changes. In contrast to current complex schemes that require the assembly of discrete sensors for multiparameter monitoring, our prototype with dual-modal capability demonstrates a significant advancement toward compact, multifunctional wearable health monitors.

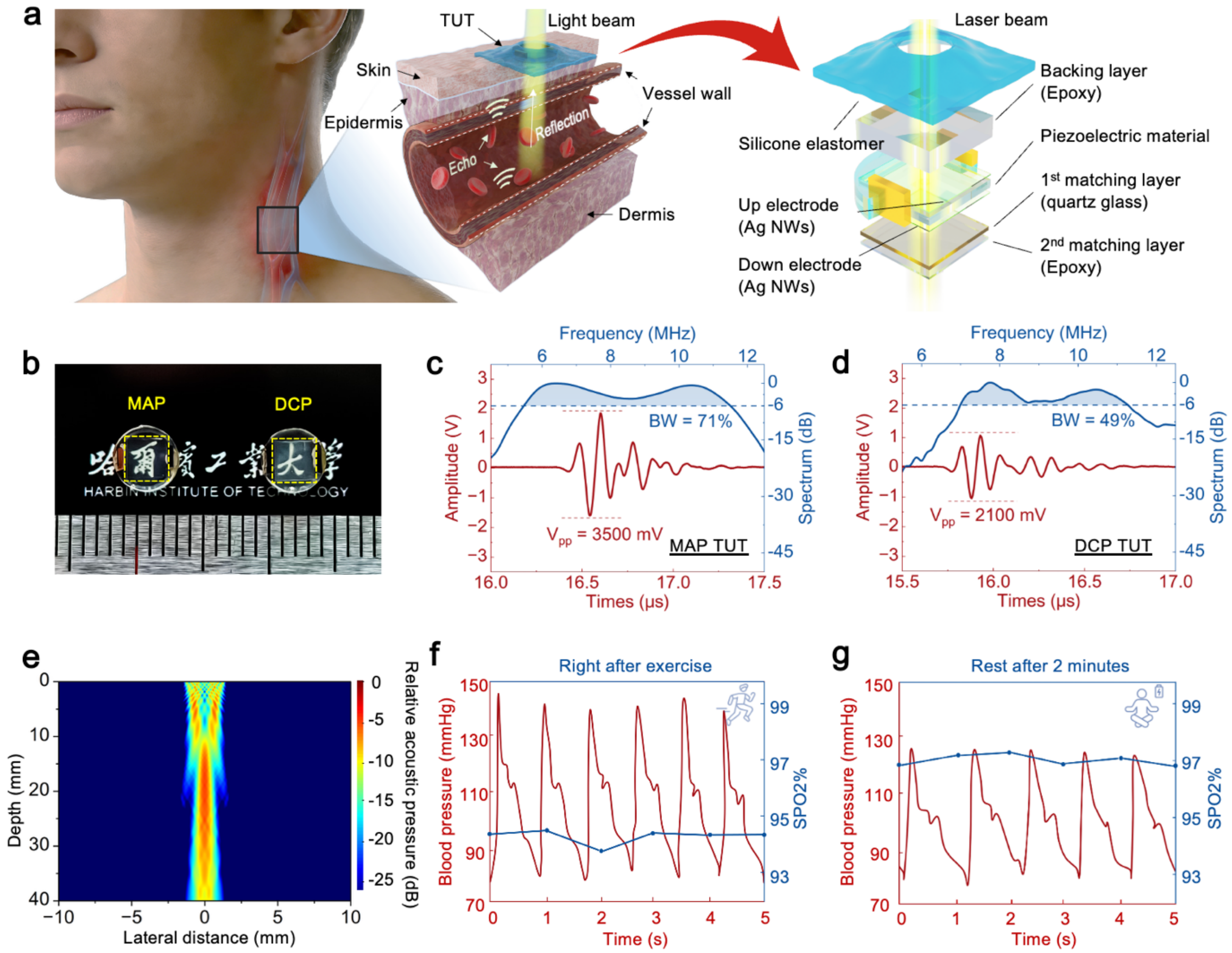

**Fig. 5 A dual-modal wearable physiological sensor enabled by transparent piezoelectricity.** **a,** Schematic of the sensor, which simultaneously monitors blood pressure via transparent ultrasound transducer (TUT) and peripheral oxygen saturation ($SpO_2$) via photoplethysmography through the TUT. **b,** Optical contrast between the transparent MAP and partially opaque DCP

piezoelectric elements, highlighting the essential optical pathway for $SpO_2$ sensing. **c,** Pulse-echo response and frequency spectrum of the MAP-based TUT with short spatial pulse length (~0.5 μs), a large peak-to-peak voltage ($V_{pp}$= 3500 mV) and a wide bandwidth (~71%), **d,** compared performance of the DCP-based device. **e,** Simulated acoustic field of the MAP-based TUT, demonstrating a collimated beam with sustained penetration depth in tissue. **f,** Continuous monitoring of blood pressure (BP) and $SpO_2$ from a subject during post-exercise recovery with systolic BP elevated at approximately 150 mmHg and $SpO_2$ of 95%. g, BP decreased to ~ 99 mmHg, while $SpO_2$ recovered to 97% following a 2-minute rest.

## Conclusions

In summary, we introduce a mechanically assisted poling (MAP) as an approach for on-demand symmetry reconstruction in ferroelectric single crystals. By navigating the Landau energy landscape with anisotropic electrical and mechanical fields, we convert phase-coexisting PMN-0.33PT into a stabilized 2R layered structures, stable up to 58°C with $d_{33}$ of 12,000 $pC\cdot N^{-1}$, while retaining an ultrahigh room temperature response ($d_{33}$ > 5,000 $pC\cdot N^{-1}$) and near-theoretical optical transparency (~65% in the visible range). In this framework, mechanical force acts as an external symmetry-breaking field, enabling access to thermodynamic states unattainable through conventional domain engineering. The generality of MAP is demonstrated in $BaTiO_3$, PZN-0.065PT, and PIN-0.38PMN-0.33PT crystals. The resulting material platform enables functional integration, demonstrated by a prototype of dual-modality physiological monitor combining optical and pressure sensing for wearables. More broadly, MAP provides a route toward architected crystalline states with coupled electromechanical and optical functionalities.

**Acknowledgments:** The authors acknowledge the staffs of BL02U2 at Shanghai Synchrotron Radiation Facility (SSRF) for technical support, and are grateful to Prof. Yuefeng Nie from Nanjing University for valuable discussions.

**Funding:**

Key Program of National Natural Science Foundation of China 52192661 (HL)

National Key Research and Development Program of China 2021YFF0501001 (XL)

National Natural Science Foundation of China 12372148 (XL)

National Natural Science Foundation of China 52322803 (XX)

Beijing Municipal Natural Science Foundation JQ24011 (QL)

**Author contributions:**

Conceptualization: JF, CW, XL, HL

Methodology: JF, CW, XL, YM, ZH, YQ, YD, XZ, CY, YZ, XZ, XLL, QL, JD

Investigation: JF, XL

Visualization: JF, XL, SC, DD

Funding acquisition: HL, XL, XX, QL

Project administration: HL, XL

Supervision: HL, XL

Writing – original draft: JF, XL

Writing – review & editing: XL, DD, XX, WC, HL

**Competing interests:** A Chinese patent entitled "A Method for Poling Domain Structures in Relaxor Ferroelectric Single Crystals with Combined Transparency and High Piezoelectric Performance" has been submitted with inventors HL, JF, and XL on behalf of HIT.

**Data, code, and materials availability:** All data are available in the main text or the supplementary materials.

**Supplementary Materials**

Materials and Methods

Figs. S1 to S14

Tables S1 to S3

References (*42-51*)

# Supplementary Materials for

## Mechanically Assisted Symmetry Reconstruction for Extraordinary Piezoelectricity

Jinhui Fan†, Chonghe Wang†, Xiaoyan Lu*, Yunpeng Ma, Zijian Hong, Yuzhao Qi, Yanzhe Dong, Xiaoyue Zhang, Chuchu Yang, Yongchun Zou, Xu Zheng, Xiaolong Li, Qian Li, Xiang Xu, Si-young Choi, Jiyan Dai, Wenwu Cao, Dragan Damjanovic*, Hui Li*

†These authors contributed equally to this work.
*Corresponding authors. Emails: luxy@hit.edu.cn; dragan.damjanovic@epfl.ch; lihui@hit.edu.cn

**The PDF file includes:**

## Materials and Methods

Processing of ferroelectric single crystals

To achieve both high transparency and piezoelectricity in PMN-*x*PT crystals, we chose compositions in the MPB region. The PT content was determined using the recommended formula: $x = (T_d + 58.8\ °C \pm 1.8\ °C)/631\ °C$ with $x$ the molar fraction of $PbTiO_3$ and $T_d$ the depolarization temperature (*42*). The PMN-*x*PT crystals were grown by modified Bridgman method, and were oriented by an X-ray diffraction (XRD) method with the *x*, *y* and *z* axes along the pseudocubic (pc) $[100]_{pc}$, $[010]_{pc}$ and $[001]_{pc}$ directions, respectively. Gold electrodes were sputtered onto both $(001)_{pc}$ faces of samples for different poling processes. To reduce the impact of internal stress during crystal polishing and sputtering, all crystals were annealed at 350 °C for 2 hours before the poling process. Selected samples with sample size of $4.2\times 3.2 \times 0.5\ mm^3$ were cut into $[110]_{pc}$, $[1\bar{1}0]_{pc}$, and $[001]_{pc}$ directions to apply mechanical force along $[1\bar{1}0]$ direction. The difference in lateral dimensions was used to distinguish the section for force loading.

The DC and MAP processes were performed with a function generator (Agilent 33220A, USA) in connection with a Trek 610E high-voltage amplifier (Trek Inc., USA). For the DC process, the electric field was raised at a rate of $0.2\ kV\cdot cm^{-1}\cdot s^{-1}$ and kept at $14\ kV\cdot cm^{-1}$ for 5 minutes along the $[001]_{pc}$ direction. For the MAP process, mechanical force was applied along $[1\bar{1}0]_{pc}$ direction using a homemade linear slider equipped with a micro sensor (SBT641C, Simbatouch, China) for quantification. The mechanical force on the sample is converted to an equivalent pressure $F/A$, ranging from 100 MPa to 140 MPa (depending on the sample composition), where $F$ is the force obtained from the sensor and $A$ is the area of the $(1\bar{1}0)_{pc}$ surface. Before all measurements, samples were aged for more than 24 hours after poling at room temperature.

PLM observations and RSM measurements

$[001]_{pc}$-oriented PMN-0.33PT crystals were processed as follows: $(001)_{pc}$ surfaces were mirror-polished using Buehler MasterMet 2 non-crystalizing colloidal silica polishing suspension (20 nm). Gold electrodes were sputtered onto both $(001)_{pc}$ surfaces to enable DC and MAP treatments, and subsequently chemically removed using a KI-$I_2$ solution (KI: $I_2$: $H_2O$ = 8:1:100) for further measurements.

PFM, SHG microscopy studies

All samples were cut into $0.5//[001]_{pc}\times0.5//[110]_{pc}\times3.2//[1\bar{1}0]_{pc}\ mm^3$ bars for domain observation on the $(110)_{pc}$ surface. DC electric field of $14\ kV\ cm^{-1}$ was applied along $[001]_{pc}$ at room temperature for 5 min to establish initial polarization alignment. Concurrently, uniaxial compressive stress (100 MPa) was imposed along the $[1\bar{1}0]_{pc}$ direction via a homemade linear slider. After the processing, $(110)_{pc}$ surfaces were polished using 20 nm colloidal silica suspension for PFM studies. To enable domain structure visualization via optical microscopy, mirror-polished samples undergo preferential etching in dilute hydrofluoric acid (HF) solution (the mass ratio of HF: $H_2O$ was 1:4) (*43*). The acid-etched samples with different poling processes (DC and MAP) were further characterized by SHG microscopy on $(110)_{pc}$ surfaces to correlate surface topography with ferroelectric domain symmetry.

In situ observation of domain structures

Single crystals of $BaTiO_3$, PMN-0.33PT, and PIN-0.38PMN-0.33PT were oriented and cut along specific pseudocubic crystallographic directions. $BaTiO_3$ samples were prepared with dimensions of $3//[100]_{pc}\times0.3//[010]_{pc}\times0.5//\ [001]_{pc}$ mm³, enabling the application of uniaxial compressive stress along $[100]_{pc}$ and an electric field along $[001]_{pc}$. The experiments were conducted at 10 °C using a Linkam MFS350 stage (Linkam Scientific Instruments, UK), where a DC electric field of 10 kV cm$^{-1}$ and a uniaxial stress of 200 MPa were applied. This temperature was selected to stabilize the coexistence of orthorhombic (O) and tetragonal (T) phases. Similarly, PMN-0.33PT and PIN-PMN-PT crystals were cut into identical sample sizes of $3//[110]_{pc}\times0.3//[1\bar{1}0]_{pc}\times0.5//[001]_{pc}$ mm³. For PMN-0.33PT and PIN-PMN-PT crystals, uniaxial stress was applied along $[1\bar{1}0]_{pc}$ and the electric field along $[001]_{pc}$. The stress of 100 MPa was applied using the Linkam MFS350 system with an application of DC field of 14 kV cm$^{-1}$ at room temperature.

Piezoelectricity and dielectricity measurements

The piezoelectric coefficient $d_{33}$ were evaluated by a quasi-static $d_{33}$ meter (ZJ-3A+, Institute of Acoustics, China). The electric-field-induced strain was measured via ferroelectric tester system (Premier II, Radiant Technologies, USA) with a laser interferometer (SIOS SP-S 120E). The dielectric permittivity $\varepsilon_{33}$ was measured using an LCR meter (E4980A, Keysight Technologies, USA).

Optical transmittance measurements

Transmission spectra were measured with an UV-VIS spectrophotometer (UVmini-1240, Shimadzu, Japan) at wavelengths ranging from 300 to 1,000 nm. The light was incident along the $[001]_{pc}$ poling direction. According to the Fresnel equations, the reflection loss for light with wavelengths from 450 to 1000 nm at the crystal surfaces was calculated through:

$$R = \frac{(n-1)^2}{n^2+1}$$

where $n$ is the wavelength-dependent refractive index given by the modified Sellmeier equation (*44*). Given that the refractive index for relaxor ferroelectric single crystal PMN-$x$PT ($0.24 < x < 0.33$) ranges from 2.5 to 2.7 at wavelength of 500 nm, the corresponding reflection loss is calculated to be between 31.03% to 34.86%, corresponding to ~ 65% transmittance for visible light.

Polarizing light microscopy imaging

The macro domain patterns were observed on $(001)_{pc}$ surface using a 0°/90° crossed polarizer/analyzer PLM system (Zeiss Axioscope 40, Carl Zeiss, Germany). The extinction behaviors were quantified with an automatic full-angle light intensity detection (AFALID) method (*45*).

Synchrotron three-dimensional reciprocal space mapping

Three-dimensional reciprocal space mapping (3D-RSM) experiments were conducted at Beamline BL02U2 of the Shanghai Synchrotron Radiation Facility (SSRF, China) to examine the phase structures of DCP and MAP processes. Monochromatic X-rays with a wavelength of $\lambda$ = 0.7006 nm were used, and diffraction patterns were collected using an EigerX 500K pixel area detector. Specific reflections were measured via $\theta$-axis scans with angular increment of 0.01°. The $(202)_{pc}$ reflection was scanned from 14.643° to 16.243° for the O-dominant samples and from 13.653° to 15.253° for the R-dominant samples. The $(222)_{pc}$ and $(\bar{2}22)_{pc}$ reflections were scanned from 16.75° to 18.35° and from 16.78° to 18.38° for R samples, respectively. Each frame had an exposure time of 1 second. The 3D-RSM images were then reconstructed using custom-developed Python scripts.

Optical second harmonic generation measurements

Second-harmonic generation (SHG) was performed using a home-designed scanning microscope. The excitation source was a Ti: sapphire mode-locked femtosecond laser (MaiTai SP, Spectra-Physics), generating a linearly polarized beam at 800 nm with a spectral bandwidth of 30 nm. Before imaging, the sample was oriented such that its crystallographic axes aligned with microscope coordinate system: $[\bar{1}10]_{pc}$, $[001]_{pc}$, and $[110]_{pc}$ were set parallel to the $X$ (polarizer axis), $Y$ (analyzer axis), and $Z$ (optical axis) directions, respectively. This alignment ensures that the initial polarization ($\varphi$) and analyzer ($\alpha$) angles are 0° when the incident and detected polarization states are both parallel to the sample's $[\bar{1}10]_{pc}$ direction. Following the initial alignment, the polarizer and analyzer were co-rotated counterclockwise to maximize the SHG intensity (yielding $\alpha = \varphi \approx 30°$ for the R-dominant sample, and $\alpha = \varphi \approx 30°$ or 135° for the O-dominant sample). SHG imaging were performed with a scanning resolution of 0.2 μm. The scanning area was set to 30 × 30 $\mu m^2$ for the O-dominant samples and increased to 60 × 60 $\mu m^2$ for the R-dominant samples due to their larger domain size. The SHG polarimetry were collected by fixing the analyzer at $\alpha$ = 0° or 90° while rotating the polarization angle ($\varphi$) for datasets $I^{2\omega}(\varphi, 0°)$ and $I^{2\omega}(\varphi, 90°)$ which enables the distinction between the R phase (point group 3*m*) and O-phase (point group *mm*2) through their symmetry-dependent nonlinear susceptibilities (*46,47*).

Piezo-response force microscopy (PFM) studies

The PFM studies were carried out on an MFP-3D (Asylum Research) using Ir/Pt-coated conductive tips (Nanosensor, PPP-NCLPt). The PFM imaging scan uses alternating current (AC) driving voltage of 600 mV in dual AC resonance tracking (DART) mode.

Stable-polarization-state calculations

The calculated stable polarization states under electric and mechanical fields were determined by Landau-Devonshire theory. The change of elastic Gibbs free energy $\Delta G$ was calculated as follows (*48*):

$$
\begin{aligned}
\Delta G = {} & \alpha_1\left(P_1^2+P_2^2+P_3^2\right)+\alpha_{11}\left(P_1^4+P_2^4+P_3^4\right)+\alpha_{111}\left(P_1^6+P_2^6+P_3^6\right) \\
& +\alpha_{112}\left[\left(P_1^4P_2^2+P_2^4P_1^2\right)+\left(P_1^4P_3^2+P_3^4P_1^2\right)+\left(P_2^4P_3^2+P_3^4P_2^2\right)\right] \\
& +\alpha_{12}\left(P_1^2P_2^2+P_1^2P_3^2+P_2^2P_3^2\right)+\alpha_{123}P_1^2P_2^2P_3^2 \\
& -\tfrac{1}{2}s_{11}\left(\sigma_1^2+\sigma_2^2+\sigma_3^2\right)-s_{12}(\sigma_1\sigma_2+\sigma_2\sigma_3+\sigma_3\sigma_1) \\
& -\tfrac{1}{2}s_{44}\left(\sigma_4^2+\sigma_5^2+\sigma_6^2\right)-Q_{11}\left(\sigma_1P_1^2+\sigma_2P_2^2+\sigma_3P_3^2\right) \\
& -Q_{12}\left[\sigma_1\left(P_2^2+P_3^2\right)+\sigma_2\left(P_3^2+P_1^2\right)+\sigma_3\left(P_1^2+P_2^2\right)\right] \\
& -Q_{44}(\sigma_4P_2P_3+\sigma_5P_1P_3+\sigma_6P_2P_1) \\
& -E_1P_1-E_2P_2-E_3P_3
\end{aligned}
$$

where $P_i$ is the polarization component; $\sigma_j$ is the stress component in Voigt's notation; $\alpha_i$, $\alpha_{ij}$, $\alpha_{ijk}$ are Landau coefficients adapted from ref. (*49*) (see section Phase field modelling for numerical values) using the experimental ferroelectric, piezoelectric and dielectric properties of PMN-0.33PT crystals at room temperature; $s_{ij}$ is the elastic compliance coefficient at constant polarization (*50*); and $Q_{ij}$ is the electrostrictive constant (*51*). The steady state of polarization is determined by minimizing the Gibbs free energy numerically via the Newton-Raphson method (ensuring an error in $P_i$ below $10^{-12}$ C m$^{-2}$). To visualize the existing polarization variants, we project the Gibbs free energy $\Delta G(P_1, P_2, P_3)$ onto a spherical coordinate system as $\Delta G(|P_s|, \theta, \varphi)$. Polarization magnitude $|P_s|$ was computed via Newton-Raphson iteration across spherical coordinate grids ($\Delta\theta = \Delta\varphi = 1°$; $\theta \in [0°, 180°]$, $\varphi \in [0°, 360°]$), achieving convergence with residual errors below $10^{-12}$ C m$^{-2}$.

## Orientation-dependent piezoelectricity calculations

The effective longitudinal piezoelectric coefficient ($d'_{33}$) for a single crystal along an arbitrary crystallographic orientation is calculated by transforming the piezoelectric coefficient matrix (*d*-matrix) from the standard crystallographic frame to a new coordinate system aligned with the desired direction. The procedure follows the coordinate rules for the third-rank tensor, using direction cosines that define the rotation between the original and target coordinate systems. The *d*-matrix, expressed in Voigt notation, is first defined in the fundamental crystallographic setting, with its elements subsequently obtained from stable polarization-state calculations using Landau-Devonshire theory. For a rhombohedral (R) phase crystal, the polar axis (*z*) is conventionally aligned with the $[111]_{pc}$ direction; for an orthorhombic (O) phase crystal, it is aligned with $[110]_{pc}$. The rotation of the coordinate system is described by a matrix $\alpha$, constructed from the Euler angles or directly from the direction cosines relating the new axes to the crystal's principal axes. The transformed piezoelectric matrix $d'$ is obtained via the relation:

$$d' = \alpha \cdot d \cdot \beta^{-1}$$

where $\alpha$ is the 3×3 transformation matrix for the polar vector (electric field-related) and $\beta$ is the 6×6 transformation matrix for the second-rank stress or strain tensor. The resulting effective longitudinal coefficient $d'_{33}$ is extracted as the element in the third row and third column of the transformed $d'$-matrix, representing the longitudinal piezoelectric coefficient for the specific crystal orientation under study.

Phase-field simulations

The expected domain evolution was obtained by performing phase-field simulations of the spatial distribution of the ferroelectric polarization vector $\vec{P}$ which is described by the time-dependent Ginzburg-Landau equation:

$$\frac{\partial \vec{P}}{\partial t} = -L \frac{F_{total}(\vec{P})}{\delta \vec{P}}$$

where $t$ is the time, $F_{total}$ is the total free energy and $L$ is the kinetic coefficient. The total free energy contains contributions from the bulk, elastic, electric and gradient energies. The Landau coefficients adapted from ref. (*49*) for PMN-0.33PT crystal are $\alpha_1 = 0.76\times10^5(T-393.7)$ $C^{-2}m^2N$, $\alpha_{11} = -0.475\times10^8$ $C^{-4}m^6N$, $\alpha_{12} = 0.115\times10^8$ $C^{-4}m^6N$, $\alpha_{111} = 0.375\times10^9$ $C^{-6}m^{10}N$, $\alpha_{112} = 0.245\times10^9$ $C^{-6}m^{10}N$, $\alpha_{123} = 0.295\times10^9$ $C^{-6}m^{10}$, where $T$ is the temperature in Kelvin. Based on these Landau parameters, the intrinsic dielectric permittivity at 300 K was calculated to be ~16,000, which is consistent with experimental results (Fig. 4e in main text). The parameters in the elastic, electrostatic and gradient energy density were assumed to be uniform throughout the system. In the simulations, we employed three-dimensional 200 × 200 × 100 discrete grid points. The grid space in real space was $\Delta x = \Delta y = \Delta z = 1$ nm. All simulation visuals show 100×100×100 subregions of full 200×200×100 $nm^3$ domains. Periodic boundary conditions are applied along all three dimensions to mimic a bulk single crystal. Stress free mechanical boundary conditions are employed to calculate the initial configuration. Effective average strain components are calculated and then added to the system for different effective stresses.

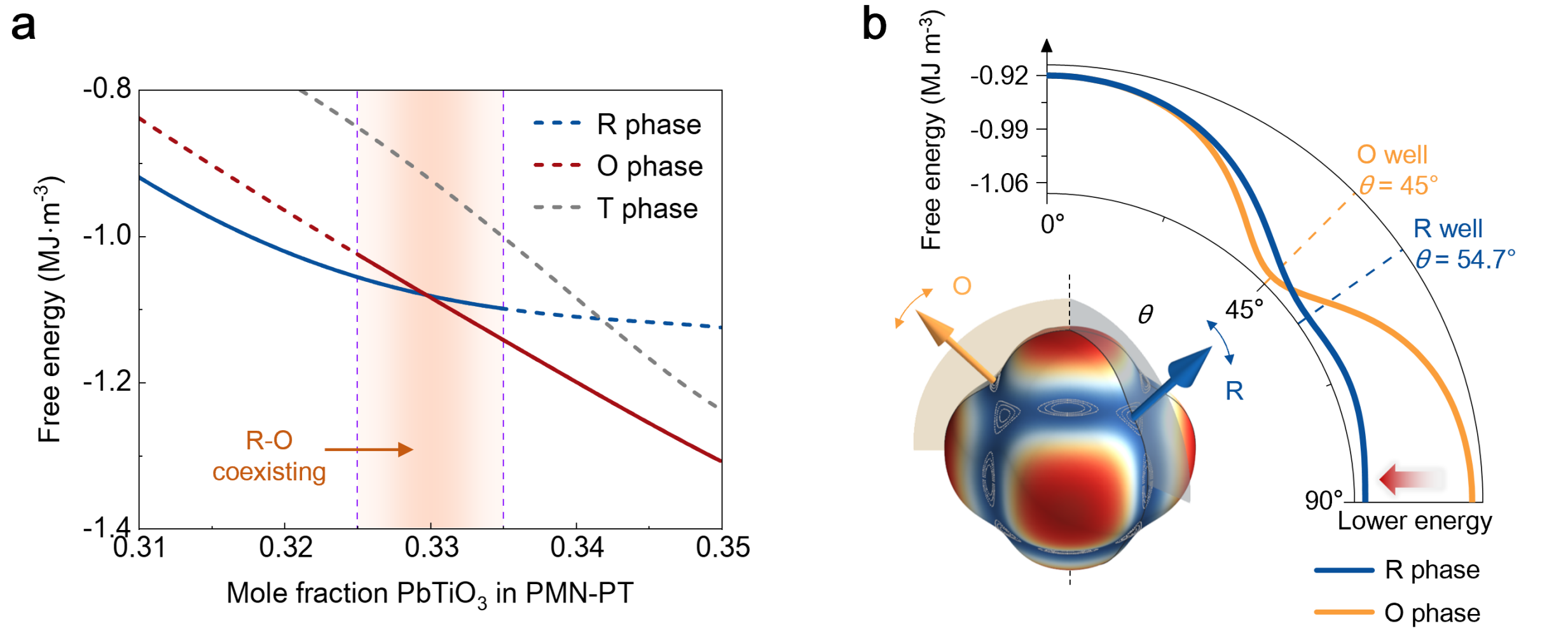

**Fig. S1.**

**Structural phase evolution and Landau free energy landscape of PMN-*x*PT near MPB. a**, Free energies as a function of composition for the R (blue), O (red), and T (grey) phases at 25 °C. Solid curves indicate the stable/metastable regions of each phase; dashed curves indicate unstable regions. The boundary between the R and O phases is near $x$ = 0.33. **b**, Spherical Landau free energy landscape $F_{\mathrm{Landau}}(\theta, \varphi)$ for PMN-0.33PT, showing nearly degenerate minima ($F_{\mathrm{Landau}}$ (O) ≈ -1.0812 × $10^6$ J $m^{-3}$， $F_{\mathrm{Landau}}$ (R) ≈ -1.0805 × $10^6$ J $m^{-3}$). The two curves (blue for R phase and orange for O phase) show the variation of Landau free energy $F_{\mathrm{Landau}}(\theta, \varphi)$ with respect to the azimuthal angle $\theta$. The R-phase potential exhibits significantly reduced curvature compared to that of the O-phase in R-O coexisting system, indicating the easier polarization rotation to R phase.

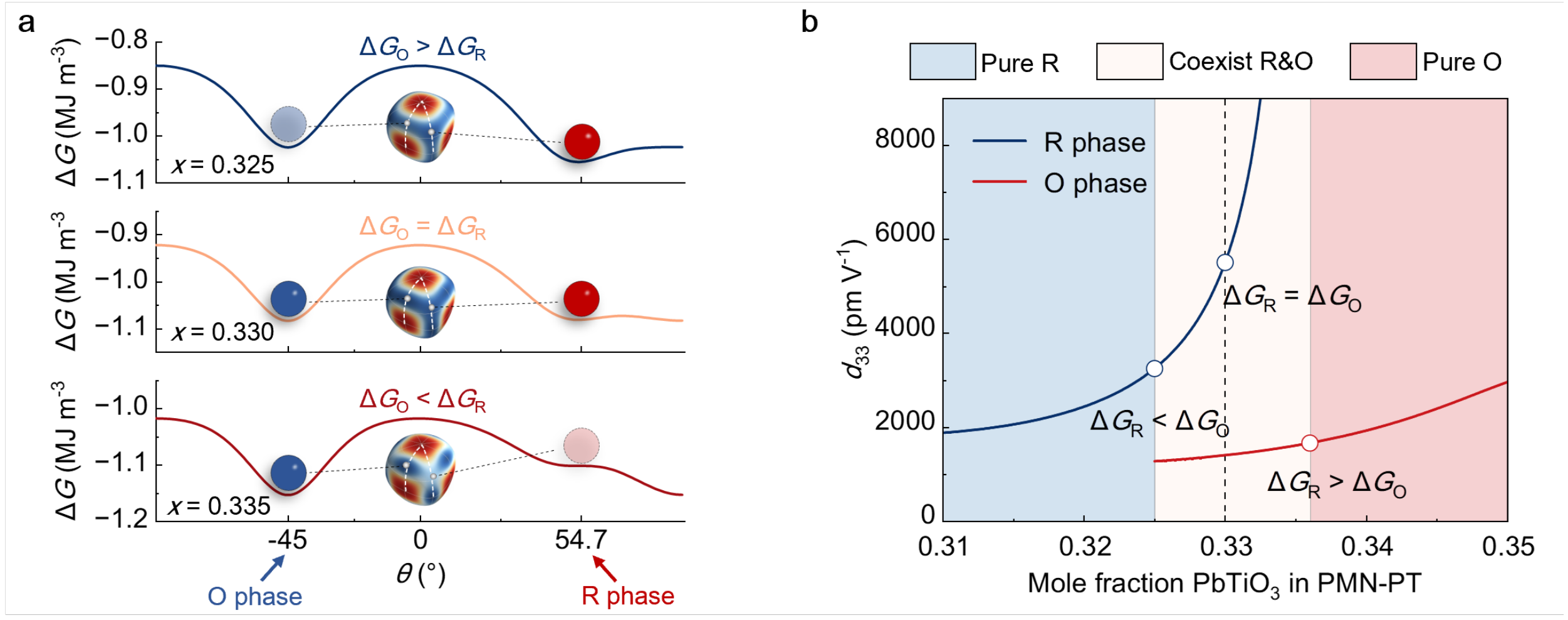

**Fig. S2.**

**Composition-driven energy landscape evolution and piezoelectric response at the MPB. a,** Free energy landscape $\Delta G(|P_s|, \theta, \varphi)$ of PMN-PT system at 300K showing R-O coexistence. At $x$ = 0.325, R phase exhibits lower energy than O phase ($\Delta G_R < \Delta G_O$). Energy degeneracy occurs at $x = 0.330$ ($\Delta G_R = \Delta G_O$), and the potential well of R phase reaches minimum depth at $x = 0.335$. **b,** Piezoelectric coefficient $d_{33}$ for pure R and O phases across MPB compositions. The value for R phase (~ 5,500 pm $V^{-1)}$ is nearly four times that of O phase (~ 1,400 pm $V^{-1}$) for PMN-PT with $x$ = 0.330. Higher $d_{33}$ values for $x > 0.330$ is not in a thermodynamic stable state.

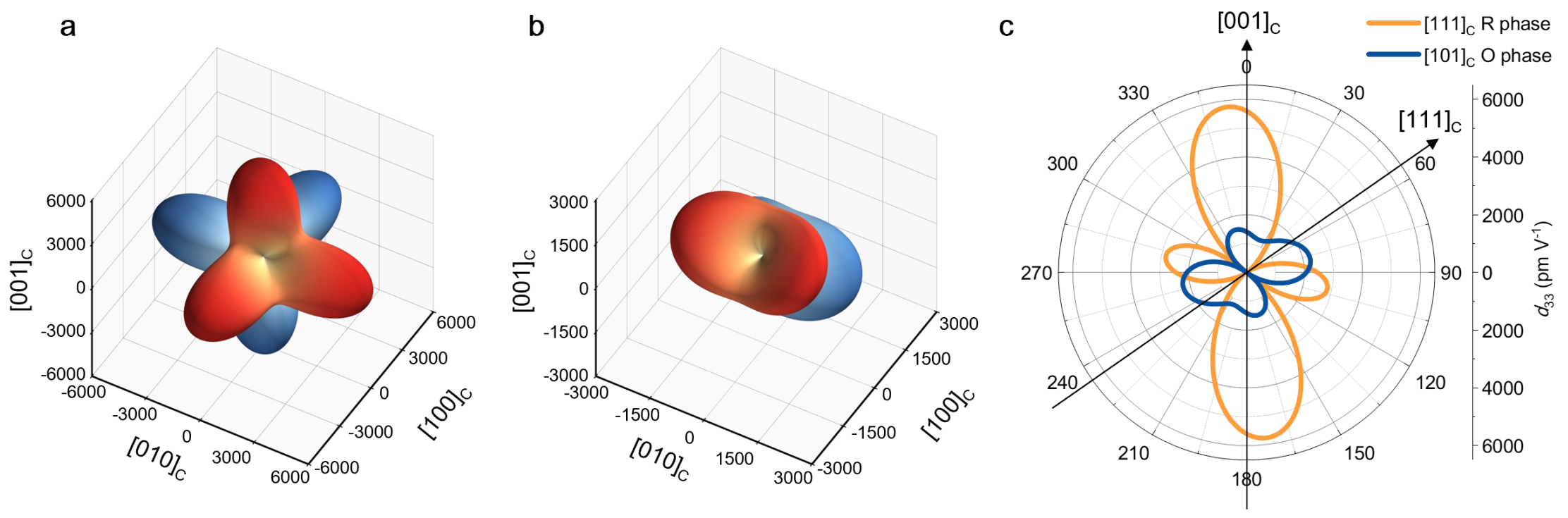

**Fig. S3.**

**Orientation-dependent piezoelectricity of pure R and O phases in PMN–0.33PT. a,** R phase. **b,** O phase. **c,** Comparison of $[001]_C$-oriented $d_{33}$. $d_{33}$ of R phase reaches approximately 5,500 pm $V^{-1}$ (orange line), a value nearly four times that of the pure O phase (~1,400 pm $V^{-1}$, blue line).

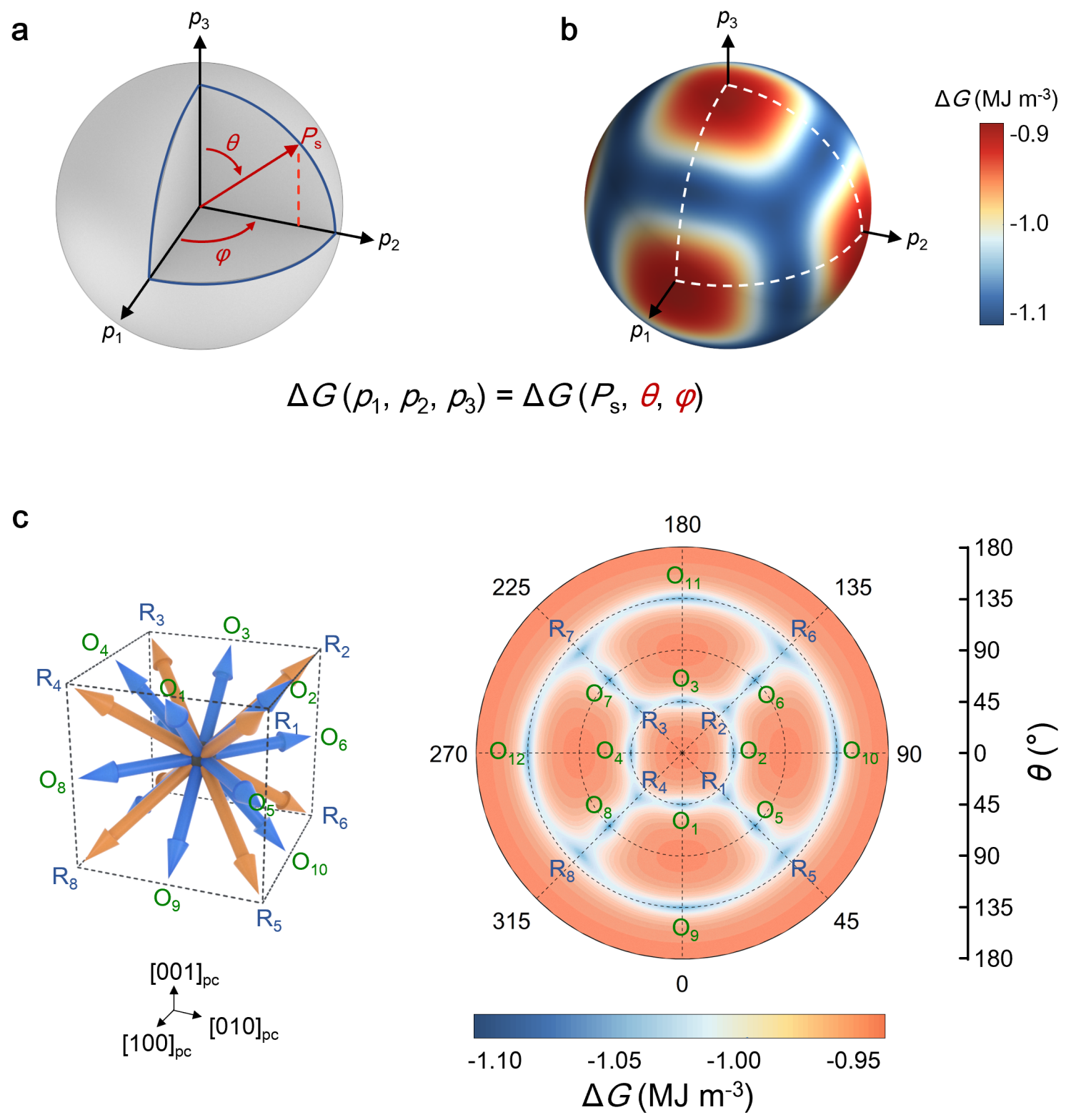

**Fig. S4.**

**Landau free energy landscape and phase coexistence in PMN-0.33PT. a,** Spherical coordinate projection of polarization vector $P_s(\theta,\ \varphi)$; **b,** Orientation-resolved free energy $\Delta G(\theta,\ \varphi)$; **c,** Degenerate polarization states in R-O phase coexisting system. **d,** Polar-coordinate Landau free energy landscape with polarization vector distributions.

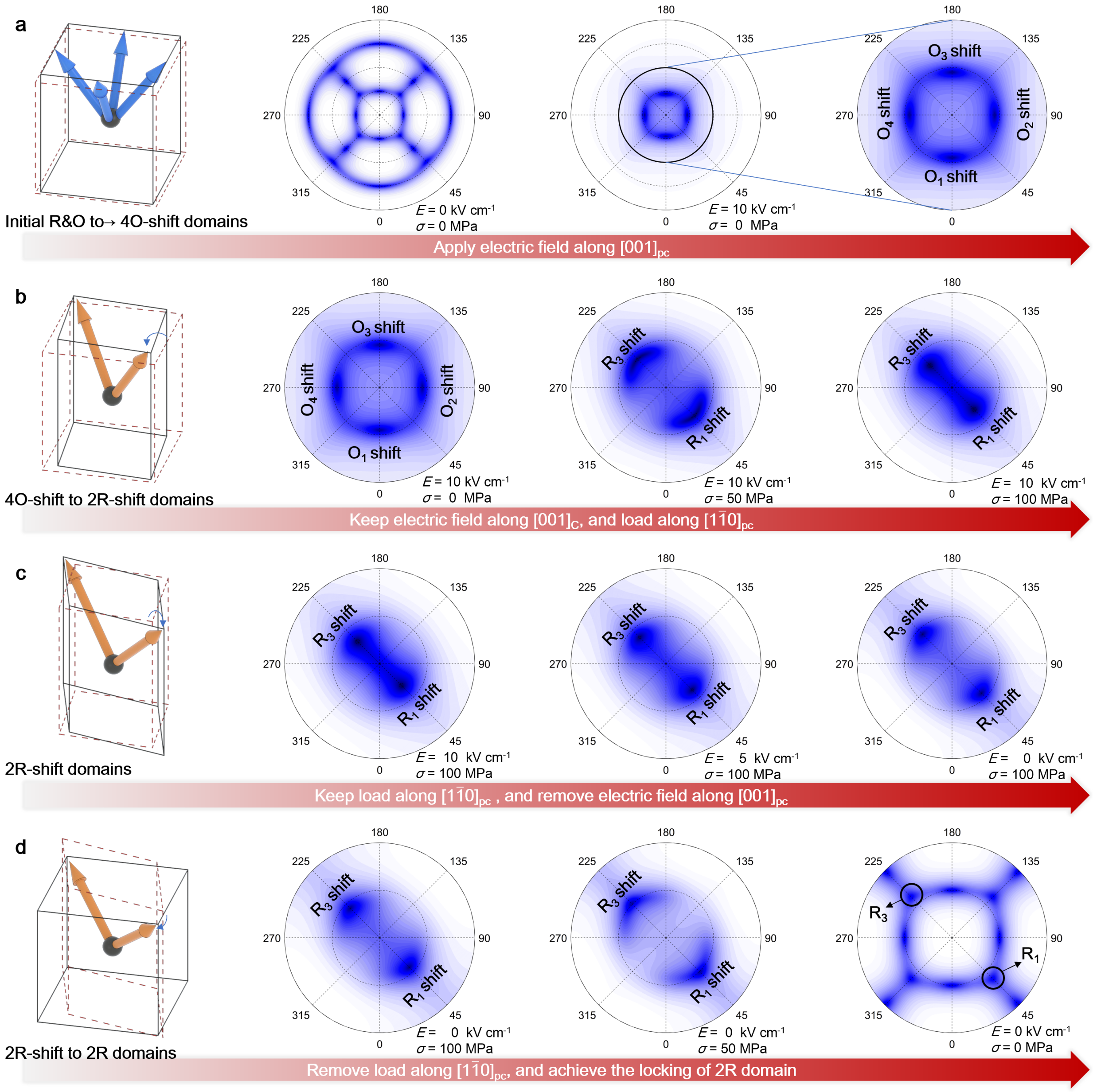

**Fig. S5.**

**MAP for 2R state from R&O coexisting system. a,** Stabilization of upward O-shift potential wells ($O_1$ - $O_4$) under electric field along $[001]_{pc}$. The obtained O states deviate slightly from ideal O phases with small shifts. **b,** Stabilization of R variants ($R_1$ and $R_3$) with further application of compressive stress along $[1\bar{1}0]_{pc}$. **c,** Preserved $R_1$ and $R_3$ variants after the removal of electric field while maintaining the mechanical loading. **d,** Preserved $R_1$ and $R_3$ variants after withdrawal of all external fields.

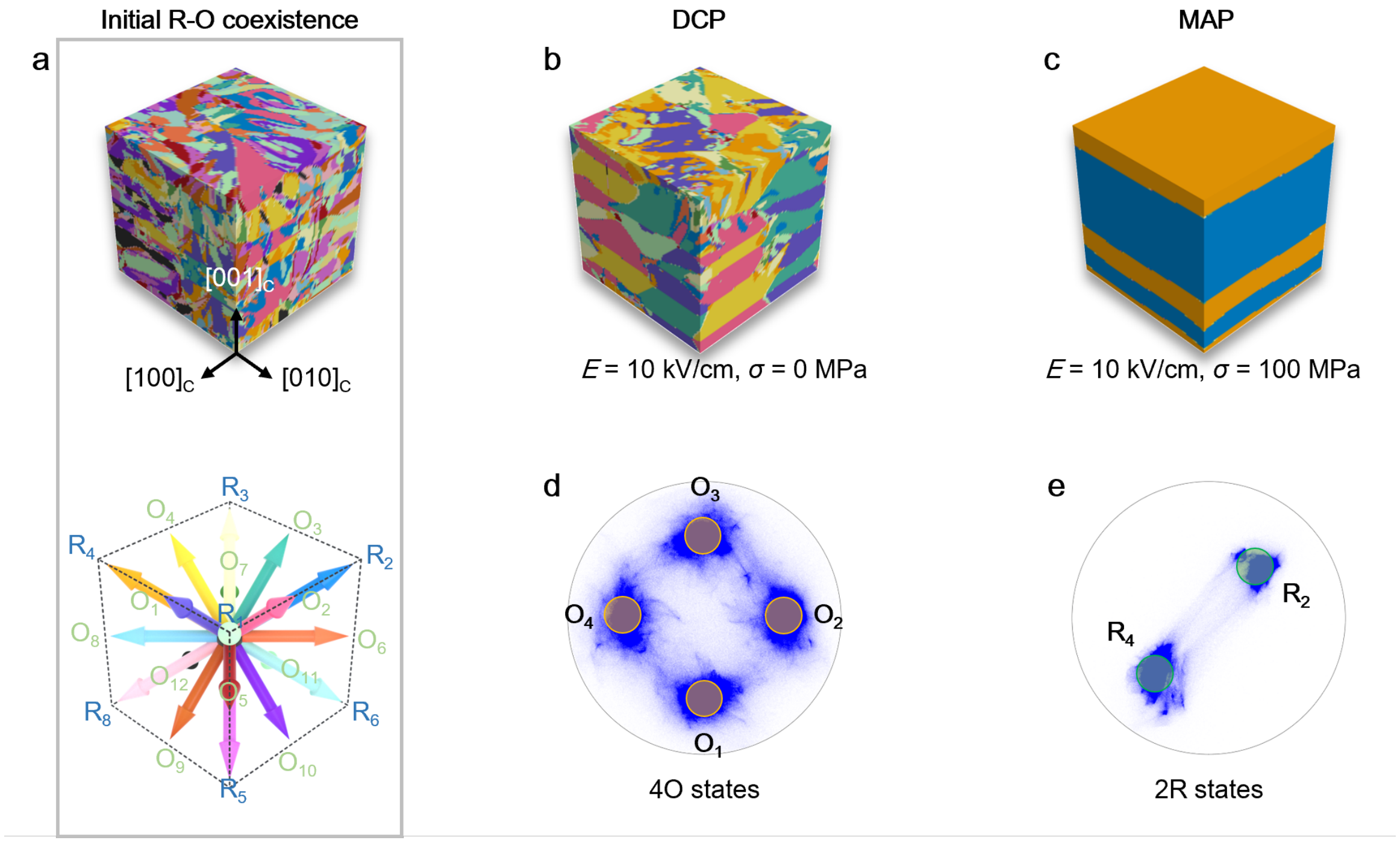

**Fig. S6.**

**Phase-field simulations of domain structures in PMN-0.33PT through DCP and MAP processes.** **a**, Unpoled state with 20 degenerate domain variants (8R and 12O). **b**, **c**, 3D domain configuration after DCP process ($E$ = 10 kV/cm with zero stress) and ACP process ($E$ = 10 kV/cm and $\sigma$ = 100 MPa). **d**, **e**, The corresponding polarization orientations showing a predominant localization near four O variants and two R-like variants, respectively.

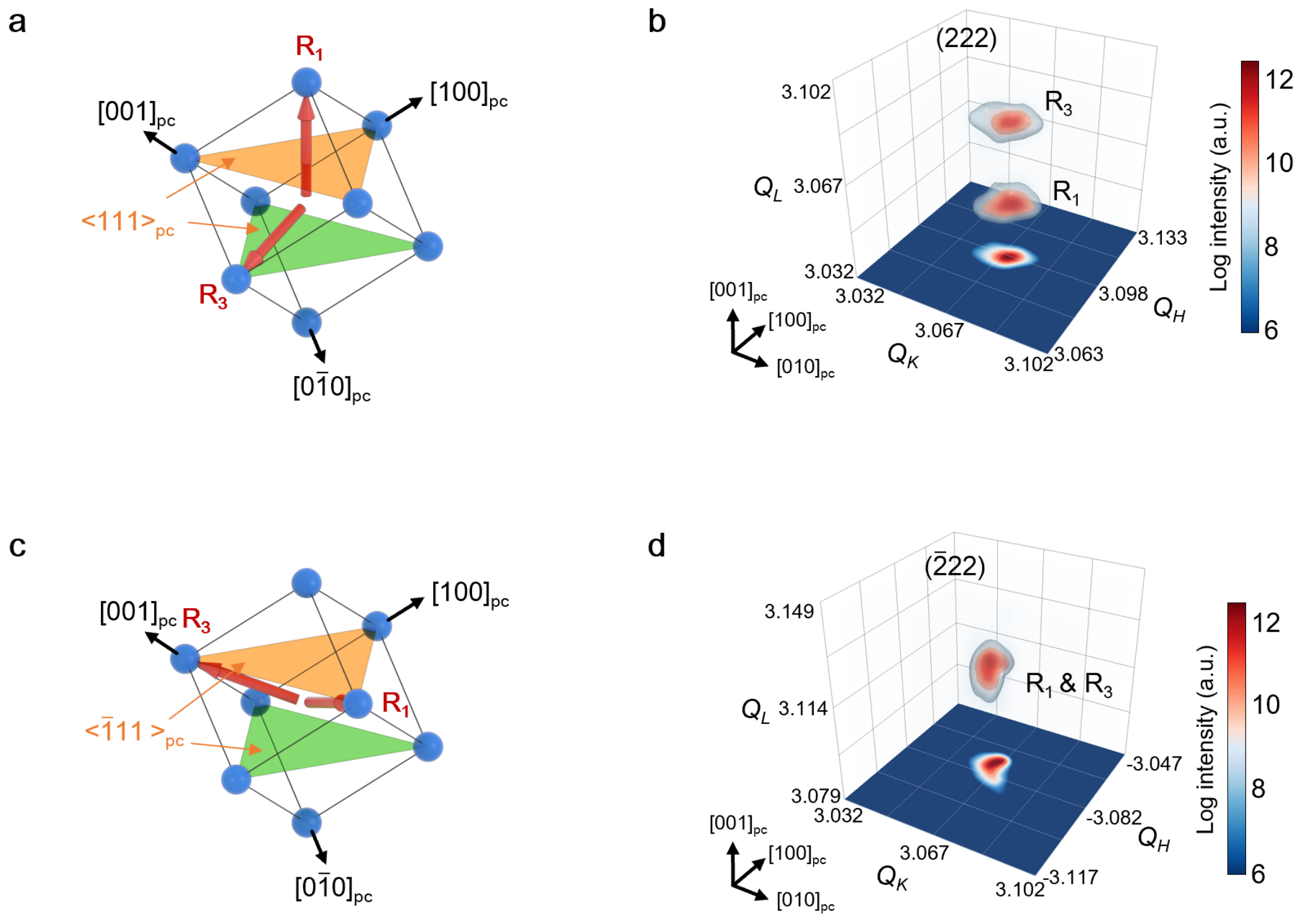

**Fig. S7.**

**Reciprocal space mapping of 2R domains in $[001]_{pc}$-oriented PMN-0.33PT crystals processed via MAP. a**, Schematic of crystal orientation and polarization variants in $<111>_{pc}$ reflection scanning. **b**, 3D-RSM of $(222)_{pc}$ reflections, displaying two distinct sets of diffraction spots attributed to the lattice distortions from $R_1$ and $R_3$ variants. **c**, Schematic of crystal orientation and polarization variants in $<\bar{1}11>_{pc}$ reflection scanning. **d**, 3D-RSM of $(\bar{2}22)_{pc}$ reflections, showing a single diffraction spot indicating equal lattice spacing of $R_1$ and $R_3$ variants in this geometry.

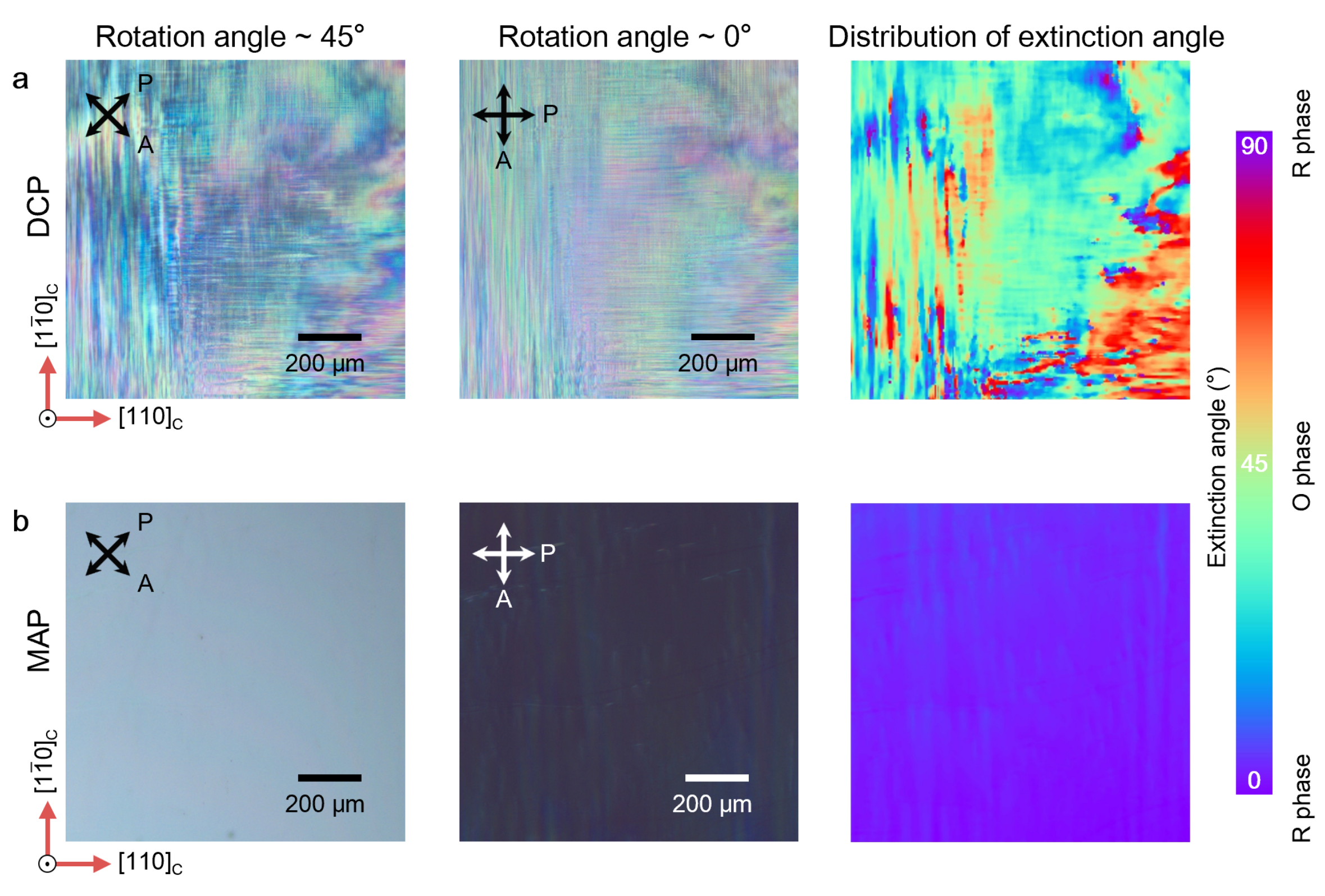

**Fig. S8.**

**Polarized light microscopy (PLM) analysis of domain structures in [001]$_{pc}$-oriented PMN-0.33PT crystals. a,** DCP-processed sample exhibiting cross-like 60° domain walls (O phase), along with non-uniform extinction angles. **b,** MAP-processed sample showing 2R domain configuration with uniform extinction angles of 0°/90°, consistent with Malus' law. The absence of visible domain walls and enhanced extinction-angle homogeneity indicate the stabilization of 2R domain structure via MAP.

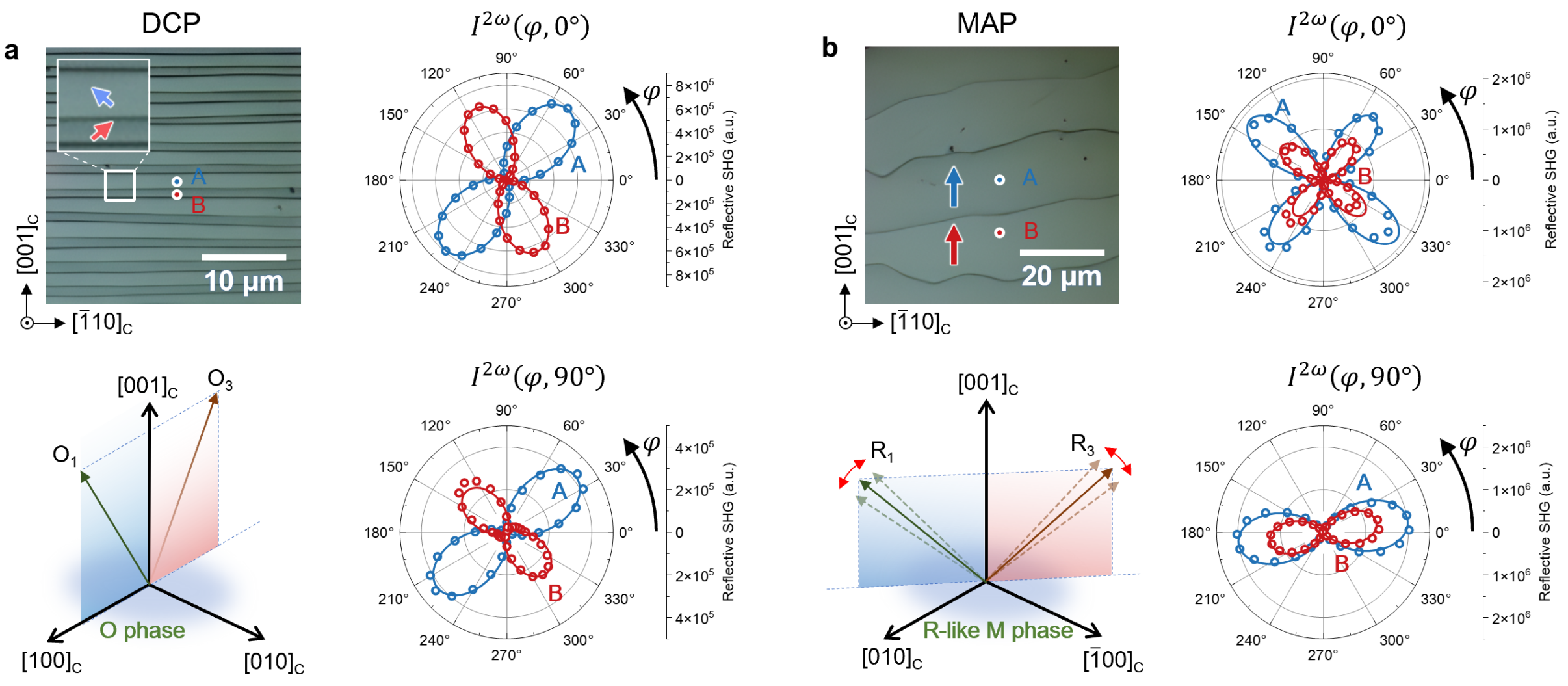

**Fig. S9.**

**Polarization-dependent second-harmonic generation (SHG) responses.** **a**, DCP-processed sample. **b**, MAP-processed sample. The DCP-processed sample exhibits polarization-resolved SHG intensities $I^{2\omega}(\varphi, 0°)$ and $I^{2\omega}(\varphi, 90°)$ with polar patterns characteristic of *mm*2 point group symmetry. Variations in polarization among adjacent O domains lead to pronounced differences in both the orientation and intensity. In contrast, the MAP-processed sample displays a clear 3*m* symmetry, with only slight out-of-plane polarization shift between adjacent R domains.

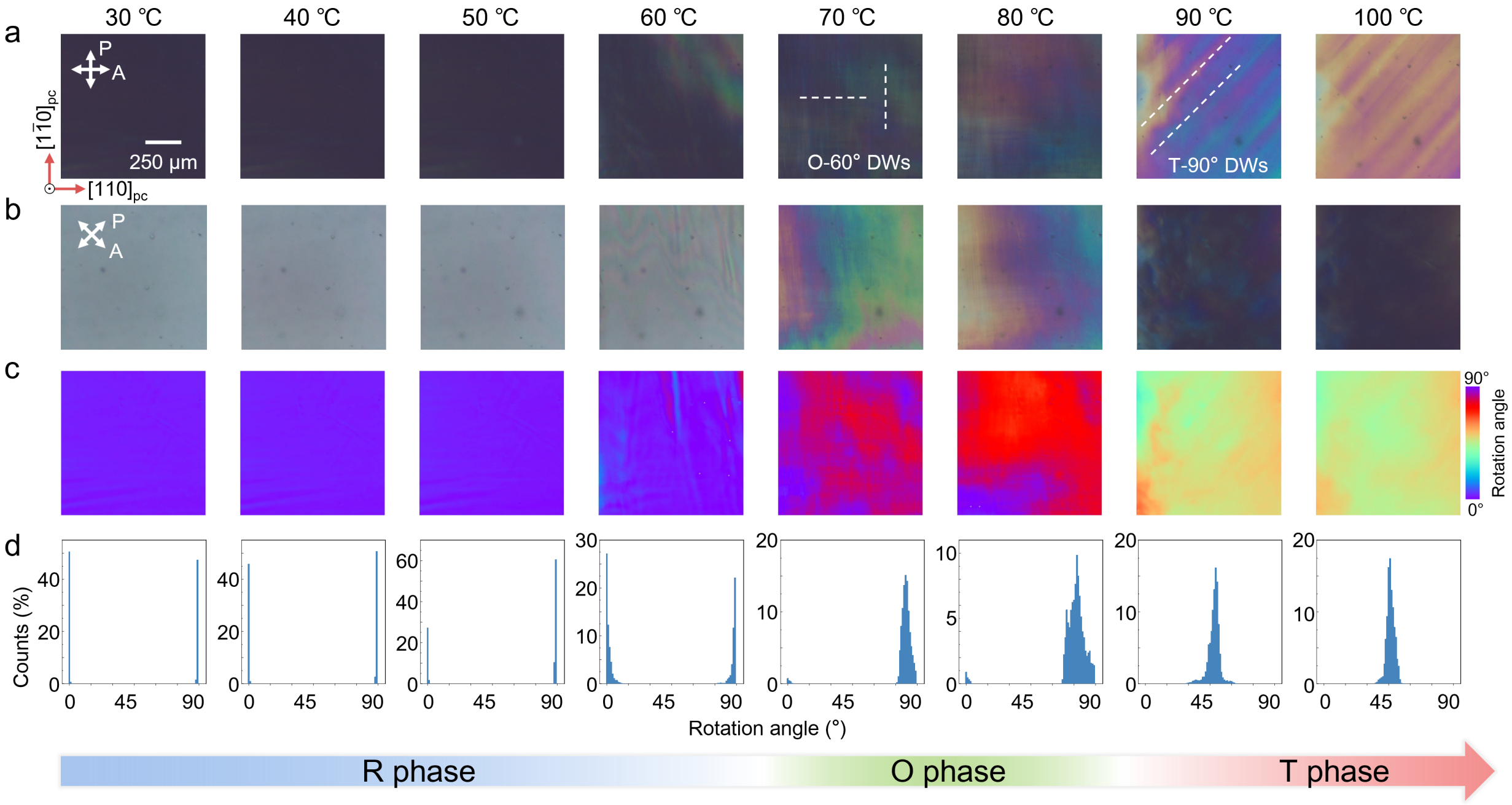

**Fig. S10.**

**Temperature-dependent domain structures of MAP-processed PMN-0.33PT crystal.** PLM images at rotation angles of **a**, 0° and **b,** 45°. Images were recorded from room temperature to 100 °C in 10 °C increments. The selected area (1 × 1 $mm^2$) is in the plane perpendicular to $[001]_{pc}$ surface. **c**, Extinction angle distributions extracted from **a** and **b,** revealing a phase transition around 60 °C. **d**, Statistical analysis of extinction angles at various temperatures. Extinction angle of the stabilized R phase (at 0° and 90°) gradually broadens with the increase of temperature and shifts toward ~75° at temperature of 60 °C, indicating the R-O phase transition. The emergence of 60° domain walls (O-60° DWs) further confirms the presence of O phase. Upon further heating, the extinction angle approaches 45°, indicating an O-T phase transition near 90 °C with 90° tetragonal domain walls (T-90° DWs).

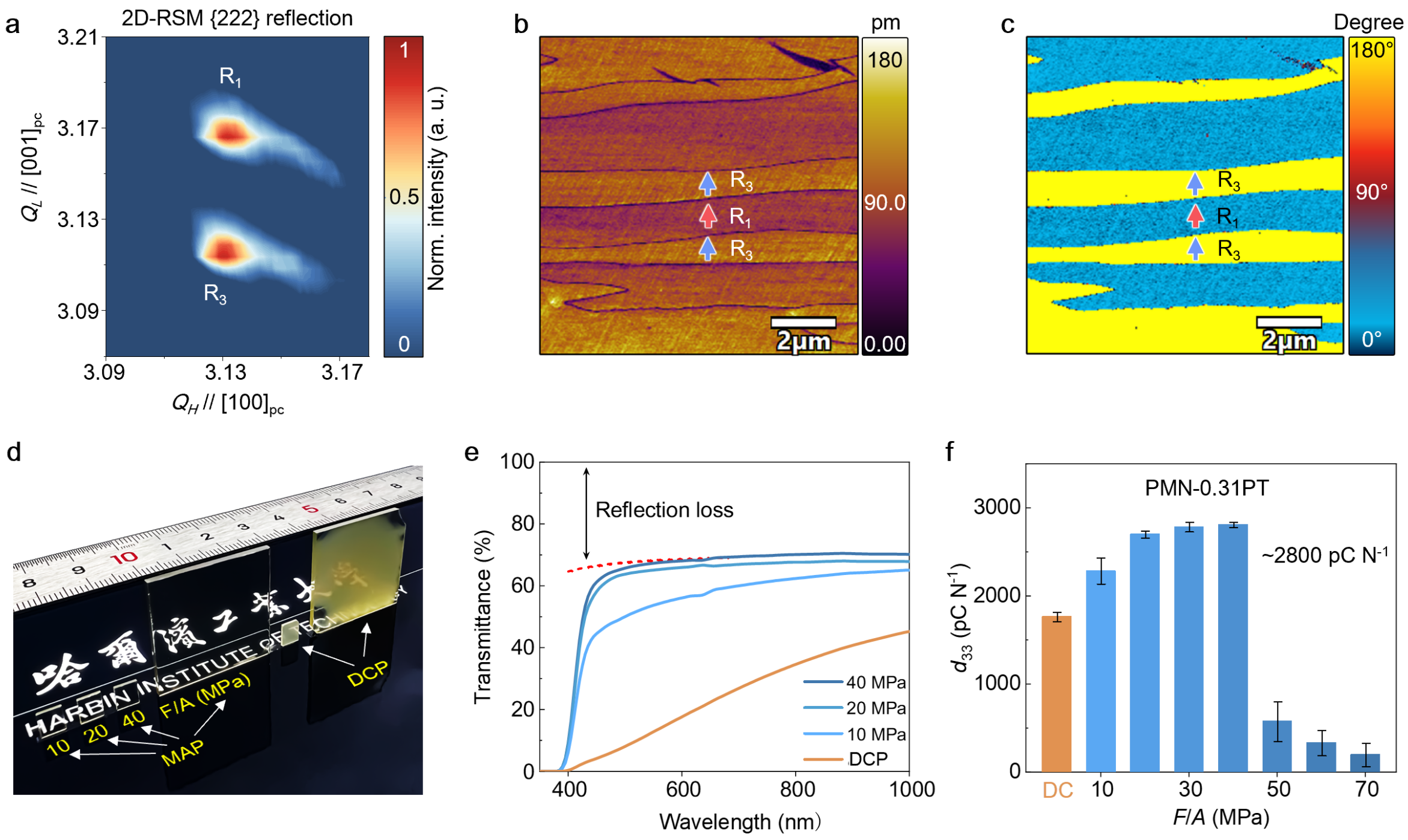

**Fig. S11.**

**Efficiency of MAP in PMN-0.31PT single crystals. a,** 2D-RSM of $(222)_{pc}$ reflection confirming 2R domain formation. **b**, **c** Phase and amplitude PFM images of the $(110)_{pc}$ plane with distinct contrast between the $R_1$ and $R_3$ variants (arrows). **d**, Optical transparency comparison of MAP-processed samples (left) and DCP-processed samples (right). The sizes of the four small samples are $4 \times 3 \times 0.5$ mm$^3$, and $20 \times 20 \times 1$ mm$^3$ for the two large ones. **e**, Transmittance spectra showing >65% visible-light transmission in MAP-processed samples, and <40% in DCP-processed samples. **f**, Comparison of piezoelectric coefficients $d_{33}$ of DCP- and MAP-processed samples under various applied external stress $F/A$, where $F$ is the applied mechanical force and $A$ is the area of $(110)_{pc}$ crystal face.

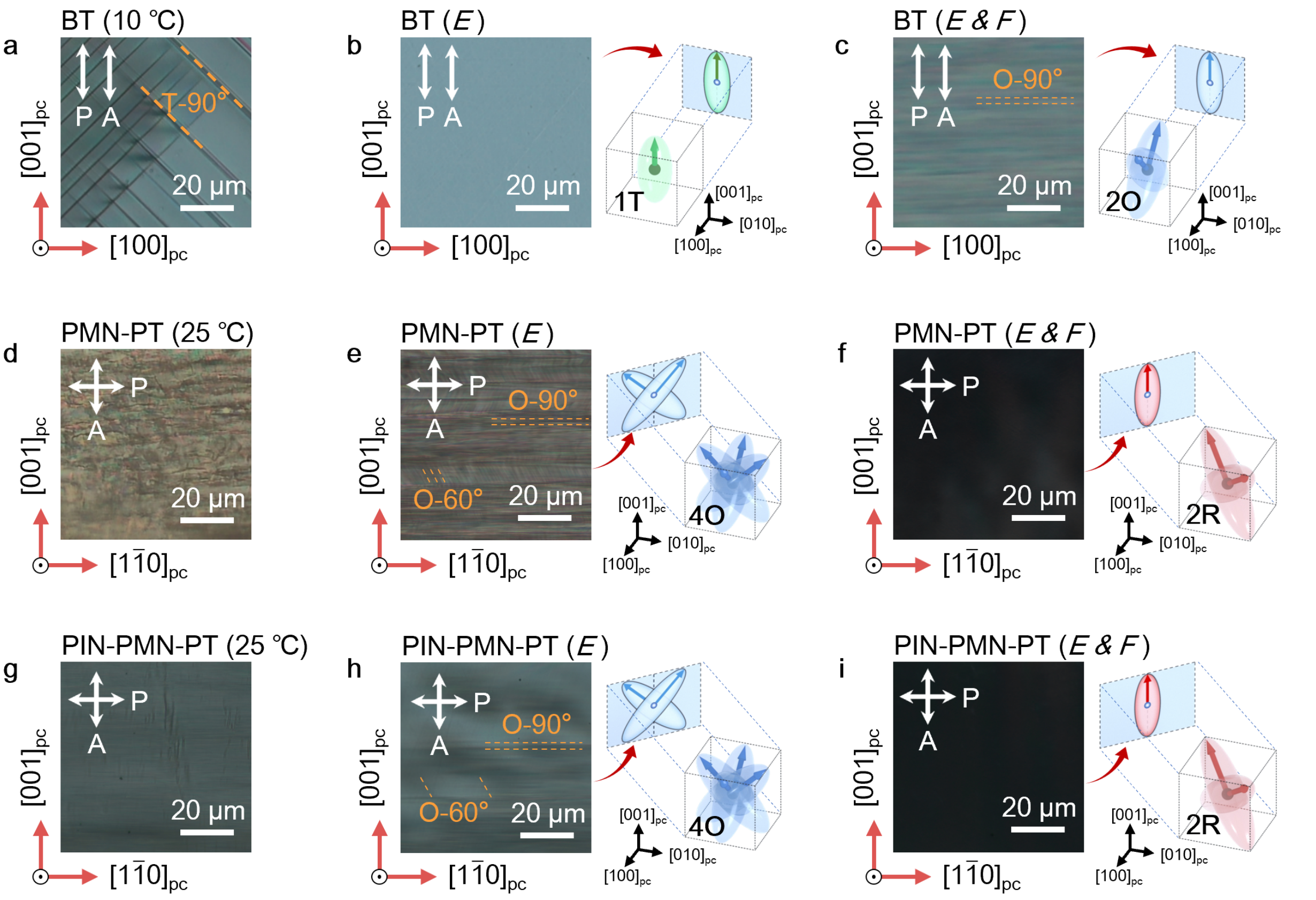

**Fig. S12.**

**In-situ PLM observations of domain structures in $BaTiO_3$ (BT), PMN-0.33PT, and PIN-0.38PMN-0.33PT single crystals under electric field and uniaxial stress. a,** BT crystal at 10 °C, showing T phase domain structure with T-90° domain walls. **b,** The crystal is poled into a monodomain 1T state with a $[001]_{pc}$ electric field (10 kV cm$^{-1}$). **c,** O phase is formed with application of uniaxial stress along $[010]_{pc}$ (200 MPa), accompanied by O-90° domain walls parallel to the $(001)_{pc}$ plane. PLM was performed with parallel polarizers due to similar optical indicatrix projections of 1T and 2O structures on the $(100)_{pc}$ plane. **d,** PMN-0.33PT crystal at room temperature with mixed-phase structure. **e,** 4O muti-domain state after a $[001]_{pc}$ electric field (14 kV cm$^{-1}$), containing O-60° and O-90° domain walls visible under crossed-PLM mode. **f,** Reconstructed 2R structure under additional stress field along $[1\bar{1}0]_{pc}$ (100 MPa). **g-i,**–PIN-0.38PMN-0.33PT crystal under the same sequence of stimuli as in d–f, showing a similar transformation from a mixed-phase state to 4O muti-domain state under electric field, and finally to 2R domain state with application of stress field. Optical extinction also exists under cross-PLM mode.

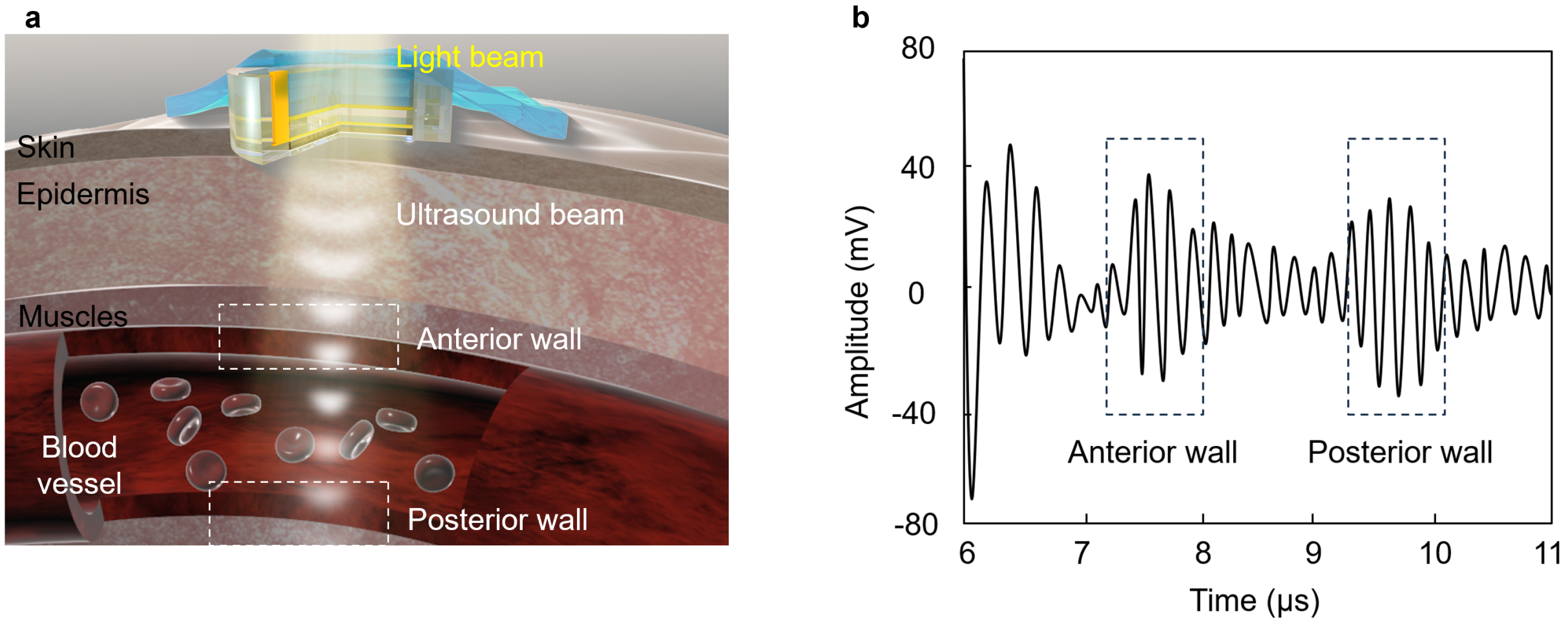

**Fig. S13.**

**Ultrasonic signal propagation and echo detection mechanism for blood pressure waveform monitoring. a,** Schematic illustration of the conformal ultrasonic patch emitting acoustic waves into multi-layered tissue structures. The ultrasound beam penetrates sequentially through the epidermis, subcutaneous tissue, and muscle layers before reaching the target blood vessel. Echoes are generated at each tissue interface due to acoustic impedance mismatches, with the most prominent reflections arising from the anterior and posterior walls of the pulsating artery. The time-of-flight (TOF) of these echoes is used to dynamically track vessel diameter changes, which are then translated into blood pressure waveforms via a calibrated viscoelastic model. **b,** Representative raw A-mode echo signals acquired from a radial artery using the wearable ultrasonic patch. The table summarizes the recorded signal amplitudes (in mV) at key tissue interfaces across successive time points (in μs). Distinct echo peaks correspond to reflections from the skin surface, muscle fascia, anterior vessel wall, and posterior vessel wall. The periodic shift in the posterior wall echo timing reflects arterial distention during the cardiac cycle, enabling real-time diameter tracking.

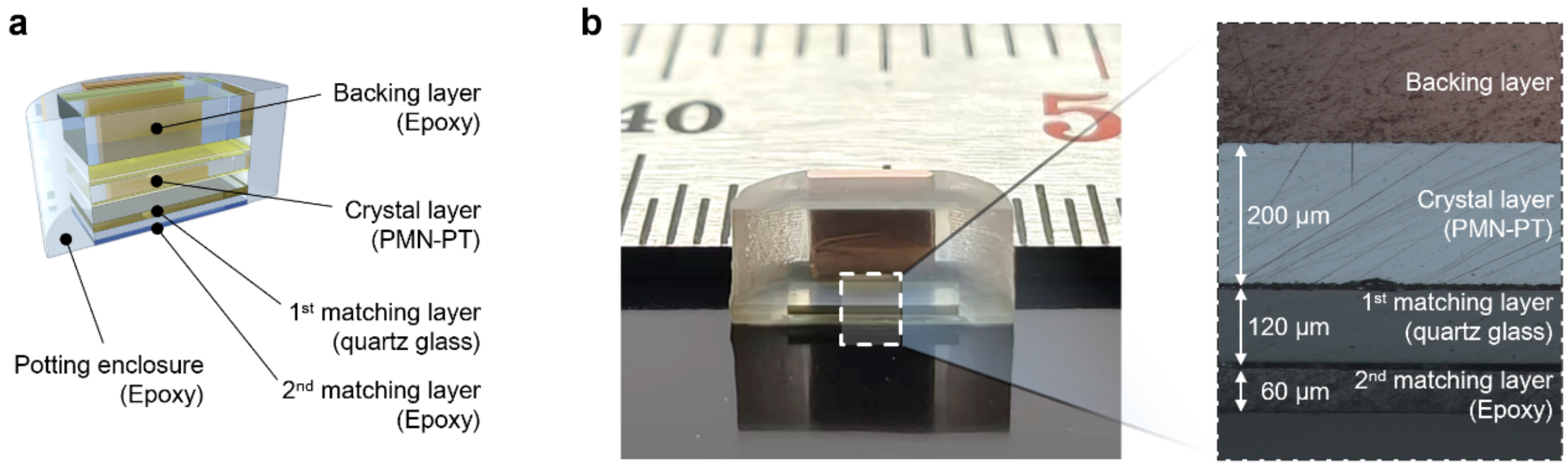

**Fig. S14.**

**Structure of PMN-PT-based transparent ultrasonic transducer (TUT). a,** Schematic diagram illustrating the layered architecture of the TUT. From bottom to top, the structure comprises: a backing layer (epoxy), a PMN-PT crystal layer, a first matching layer (quartz glass), the second matching layer (epoxy), and the potting enclosure (epoxy). **b,** Photograph of a PMN-PT-based TUT cut into half. The cross-sectional view on the right details the transducer stack with thicknesses of 200 µm for the crystal layer, 120 µm for the first matching layer, and 60 µm for the second matching layer.

| *Crystalline* | *Materials* | *Piezoelectric properties* | | *Optical Transmittance* | |
|---|---|---|---|---|---|
| | | $d_{33}$ (pC·N$^{-1}$) | References | Transmittance @500 nm (%) | References |
| | PMN-0.33PT (2R) | ~ 5100 | ***This work*** | ~ 62.5 | ***This work*** |
| | PMN-0.33PT (4O) | ~ 1600 | | ~ 40.1 | |
| | PMN-0.33PT (2O) | ~ 1600 | | ~ 63.3 | |
| | PMN-0.31PT (2R) | ~ 2800 | | ~ 65.7 | |
| | PZN-0.065PT (2R) | ~3800 | | ~62.1 | |
| | PIN-0.38PMN-0.33PT (2R) | ~4000 | | ~61.8 | |
| **Crystals** | LN | ~ 40 | *Nat. Commun.* ***15****, 10580, (2024)* | ~ 72.3 | *Ceram. Int.* ***48****, 11909-11914, (2022)* |
| | KDP | ~ 20 | *J. Chem. Phys.* ***9****, 16–33, (1941)* | > 82 (@340 nm) | *J. Opt. Soc. Am. A* ***50*** *(9), 865 (1960).* |
| | KNN | ~ 150 | *Cryst. Growth Des.* ***15****, 1180-1185, (2015).* | ~ 51.3 | *Adv. Opt. Mater.* ***10****, 2201721, (2022)* |
| | BT | ~ 190 | *Physical acoustics: principle and methods. New York: Academic Press; 1964 [chapter 3].* | ~ 60 | *Physical Review* ***96****, 801-802, (1954).* |
| | PMN-PT (R) | ~ 1600 | *Nature* ***577****, 350-354, (2020)* | ~ 10 | *Ceram. Int.* ***48****, 11909-11914, (2022).* |
| | PMN-PT (MPB) | ~ 2800 | *J. Appl. Phys.* ***90****, 3471-3475, (2001)* | ~ 17.1 | |
| | PMN-PT (T) | ~ 300 | *Phys. Status Solidi. B* ***253****, 1994-2000, (2016)* | ~ 58.5 | *J. Cryst. Growth* ***263****, 251-255, (2004)* |
| | PZN-PT (R) | ~ 2000 | *IEEE T. Ultrason. Ferr.* ***47****, 285-291 (2000)* | ~ 24.7 | *Tech. Phys. Lett.* ***30****, 1002-1004 (2004).* |
| | PZN-PT (MPB) | ~ 2900 | *IEEE T. Ultrason. Ferr.* ***49****, 1622-1627 (2002).* | ~ 38.4 | |
| | PZN-PT (T) | ~ 200 | *IEEE T. Ultrason. Ferr.* ***55****, 476-488, (2008).* | ~ 58.3 | |
| | PIN-PMN-PT (R) | ~ 1400 | *J. Alloys Compd.* ***553****, 267-269, (2013).* | ~ 48.5 | *J. Appl. Phys.* ***117****, 164104, (2015)* |
| | PIN-PMN-PT (MPB) | 2742 | *J. Appl. Phys.* ***106****, 074112 (2009).* | ~ 31.2 | |
| | PIN-PMN-PT (T) | ~ 530 | *J. Appl. Phys.* ***107****, 054107 (2010)* | ~ 66.1 | |
| | Sm: PMN-PT | ~ 4100 | *Science* ***364****, 264-268 (2019)* | ~ 31.5 | *Ceram. Int.* ***48****, 11909-11914, (2022)* |
| | Nd: PMN-PT | ~ 3500 | *Adv. Funct. Mater. 2201719, (2022)* | ~ 46.1 | |
| | PMN-0.30PT ([011]$_{pc}$-poled) | ~ 1000 | *Adv. Mater.* ***33****, 2103013, (2021)* | ~ 63.5 | *Adv. Mater.* ***33****, 2103013, (2021)* |
| | PMN-0.28PT (AC-poled) | ~ 2100 | *Nature* ***577****, 350-354, (2020)* | ~ 61 | *Nature* ***577****, 350-354, (2020)* |
| **Ceramics** | PLZT 9/65/35 | 0 | *Ferroelectrics* ***75****, 25-55, (1987).* | ~ 63.3 | *Ferroelectrics* ***75****, 25-55, (1987).* |
| | KNN-based | ~ 30 | *J. Alloys Compd.* ***784****, 60-67, (2019).* | ~ 35 | *J. Alloys Compd.* ***784****, 60-67, (2019).* |
| | Eu: PMN-25%PT | ~ 850 | *J. Mater. Chem. C* ***9****, 2426-2436, (2021)* | ~ 26.7 | *J. Mater. Chem. C* ***9****, 2426-2436, (2021).* |
| | Sm: PIN-PMN-PT | 905 | *ACS Appl. Mater. Interfaces* ***15****, 7053-7062, (2023)* | ~ 21 | *ACS Appl. Mater. Interfaces* ***15****, 7053-7062, (2023)* |

**Table S1.**
**List of optical and piezoelectric performances in bulk transparent piezoelectric materials.**

| *Materials* | PMN-0.33PT (DCP) | PMN-0.33PT (MAP) |
|---|---|---|
| Piezoelectric strain coefficient $d_{33}$ (pC·N$^{-1}$) | 1600 | 5000 |
| Piezoelectric voltage coefficient $g_{33}$ (V m·N$^{-1}$) | 0.039 | 0.038 |
| Free dielectric constant ($\varepsilon_{33}^{T}/\varepsilon_0$) | 4600 | 15000 |
| Clamped dielectric constant ($\varepsilon_{33}^{S}/\varepsilon_0$) | 410 | 650 |
| Dielectric loss (tan $\delta$) | 0.005 | 0.007 |
| Electromechanical coupling coefficient ($k_t$) | 0.55 | 0.60 |
| Longitude Velocity (m·s$^{-1}$) | 4400 | 4200 |

**Table S2.**
**Properties of available transparent piezoelectric materials.**

| Materials | Density ($kg \cdot m^{-3}$) | Velocity ($m \cdot s^{-1}$) | Acoustic impedance (MRayl) |
|---|---|---|---|
| Quartz glass | 2200 | 5500 | 12.1 |
| Epoxy (E51) | 1180 | 2700 | 3.19 |

**Table S3.**
**Material Properties of the matching and backing layers in TUTs.**